\documentclass[traditabstract,onecolumn,a4paper]{aa}

\usepackage{graphicx}
\usepackage{amsmath}
\usepackage{natbib}
\bibpunct{(}{)}{;}{a}{}{,} 

\usepackage{ulem}

\graphicspath{{./}{Figures/}}
\def\deg{\ifmmode^\circ\else$^\circ$\fi}

\def\simlt{\lower.5ex\hbox{$\; \buildrel < \over \sim \;$}}
\def\simgt{\lower.5ex\hbox{$\; \buildrel > \over \sim \;$}}

\begin{document}

\title{Iterative destriping and photometric calibration for Planck-HFI, polarized, multi-detector map-making}
\titlerunning{Iterative destriping and photometric calibration for Planck-HFI polarized map-making}

\author{M.~Tristram\inst{1} \and C.~Filliard\inst{1} \and O.~Perdereau\inst{1} \and S.~Plaszczynski\inst{1} \and R.~Stompor\inst{2} \and F.~Touze\inst{1} }

\institute{
	Laboratoire de l'Acc\'el\'erateur Lin\'eaire, Universit\'e Paris-Sud, CNRS/IN2P3, Orsay, France
	\and
	Laboratoire Astroparticule et Cosmologie, Universit\'e Paris-7 Denis Diderot, CNRS/IN2P3, Paris, France
}

\date{\today}
\abstract
{
We present an iterative scheme designed to recover calibrated $I$, $Q$, and $U$ maps from {\it Planck}-HFI data using the orbital dipole due to the satellite motion with respect to the Solar System frame. It combines a map reconstruction, based on a destriping technique, juxtaposed with an absolute calibration algorithm. We evaluate systematic and statistical uncertainties incurred during both these steps with the help of realistic, Planck-like simulations containing CMB, foreground components and instrumental noise, and assess the accuracy of the sky map reconstruction by considering the maps of the residuals and their spectra. In particular, we discuss destriping residuals for polarization sensitive detectors similar to those of {\it Planck}-HFI under different noise hypotheses and show that these residuals are negligible (for intensity maps) or smaller than the white noise level (for $Q$ and $U$ Stokes maps), for $\ell >50$. 
We also demonstrate that the combined level of residuals of this scheme remains comparable to those of the destriping-only case except at very low $\ell$ where residuals from the calibration appear. For all the considered noise hypotheses, the relative calibration precision is on the order of a few $10^{-4}$, with a systematic bias of the same order of magnitude.}


\maketitle

\section{Introduction}

The {\it Planck} mission\footnote{http://www.rssd.esa.int/planck}, launched on May 14 2009, is a third-generation satellite dedicated to observations of cosmic microwave background (CMB) anisotropies after COBE\footnote{http://lambda.gsfc.nasa.gov/product/cobe} and WMAP\footnote{http://wmap.gsfc.nasa.gov}. Its primary objectives are to measure the full-sky CMB anisotropies down to the cosmic variance limit reaching beyond $\ell \sim 2000$ in temperature and $\ell \sim 1500$ in $E$-mode polarization. Other scientific goals include in-depth studies of the Galactic emissions, extraction of catalogs of galaxy clusters and extragalactic sources and searches for non-Gaussianity.

Planck observes the sky in nine frequency bands using two instruments: {\it LFI} with three channels centered at $30$, $44$, and $70$~GHz and {\it HFI} with six channels at $100$, $143$, $217$, $353$, $545$, and $857$~GHz with an angular resolution in the HFI CMB-dominated bands, $217$ and $143$~GHz, of from $5$ to $10$~arcmin. A map-making procedure designed to recover the sky maps from the noisy time-ordered measurements is a necessary step down the data analysis pipeline of the Planck data needed to ensure a delivery of high quality sky images in the observed bands.

Map-making algorithms have been extensively tested and compared with each other within the Planck framework \citep[see e.g.][]{ashdown07a,ashdown07b} assuming some standardized, and unavoidably, idealized circumstances. This study complements these earlier works describing an iterative procedure to combine map-making and CMB orbital-dipole-based calibration procedures and investigating in detail its precision and its dependence on low-frequency correlated noise due to drifts and artifacts present in {\it HFI}-like timelines. The map-making technique used in this paper is destriping \citep[e.g.,][]{burigana99, delabrouille98, maino99, revenu00, maino02, keihanen04}, which is designed to mitigate the long-term correlation of the {\it HFI} bolometers and approach the white noise limit in the map domain. The algorithm first projects the time-ordered data (TOD) for each stable pointing period of the Planck satellite separately onto the sky resulting in a set of overlapping {\it ring}-like maps. The relative offsets of these (one per ring), i.e. ones that lead to a coherent full-sky map, are then derived using a maximum-likelihood method that marginalizes over the sky signal. The sky maps are produced by projecting the offset-corrected TOD onto the sky. Calibration is done iteratively using offsets determined from a previous step to constrain the next approximation of the calibration coefficients using the maximum likelihood algorithm. We also address the validity of the destriping hypothesis for different models of Fourier noise spectrum and estimate the effect of iteration on the calibration for the high-precision polarized map-making of Planck.

This paper is organized as follows. We first describe the algorithm for destriping and calibration in Sects.~\ref{sec:mapmaking} and \ref{sec:calibration}. Section~\ref{sec:polkapix} is devoted to {\sc Polkapix}, the implementation of those algorithms (ring-making, offset determination, calibration and projection) within the Planck-{\it HFI} Data Processing Center infrastructure. Simulations are detailed in Sect.~\ref{sec:simulations}. The results of destriping and calibration are described in Sects.~\ref{sec:destriping_results} and~\ref{sec:calibration_results}, respectively. Finally, the results of the combined method solving for offsets and gain by iteration are discussed in Sect.~\ref{sec:iteration_results}.

\section{Map-making}
\label{sec:mapmaking}

\subsection{The map-making problem}
Planck-HFI bolometers measure the brightness of the sky in a given direction convolved with an instrumental beam. The time-ordered data vector, $\mathbf{d}$, may therefore be modeled as the sum of the signal from the convolved sky $\mathbf{T}$ and the noise $\mathbf{n}$
\begin{equation}
\mathbf{d} = \mathbf{A} \cdot \mathbf{T} + \mathbf{n}.
\label{eqn:dataModel}
\end{equation}
The pointing matrix $\mathbf{A}$, of size $N_s \times N_p$, relates each time sample $s$ to the corresponding pixel $p$ in the sky. It can typically be represented as very sparse. For an axisymmetric beam response, the ``smearing" and ``pointing" operations commute, and one can solve directly for the beam-convolved map, assuming that the matrix $\mathbf{A}$ contains only one (three for polarization-sensitive detectors) non-null values in each row, as each sample is sensitive to only one pixel of the convolved sky. The data model, Eq.~\ref{eqn:dataModel} can then be written as
\begin{equation}
d_t = I_p + \varepsilon Q_p cos(2\psi_t) + \varepsilon U_p sin(2\psi_t) + n_t,
\label{eq:polDataModel}
\end{equation}
where $(I_p,Q_p,U_p)$ are beam-smoothed Stokes parameters in pixel $p$, $\varepsilon$ is the polarization efficiency (equal to zero for the total power experiments), and $\psi_t$ is the angle of the detector's polarization direction, with respect to the polarization basis in the pixel $p$, at the time $t$.

For arbitrary beam patterns, one may either need to deal with non-trivial, position-dependent smoothing in the final maps, while retaining the simplicity of the matrix $\mathbf{A}$, or to try to mitigate these effects with the help of one of the proposed beam-deconvolution techniques \citep[e.g.][]{arnau00,burigana03,armitage04}, which typically require more complex pointing matrices.

After pre-processing of the TOD, noise can typically be considered as Gaussian and piece-wise stationary so that all the statistical information of the noise is contained in its covariance matrix $\mathbf{N} = \left< \mathbf{n} \mathbf{n}^T \right>$ for which each stationary block is a symmetric Toeplitz matrix \citep{golub96}. Nevertheless, the instrumental noise is usually not white and contains some low frequency components that, if not accounted for properly, are projected to the sky as large stripes following the scan pattern.

\subsection{Maximum likelihood (ML) methods}

The most general solution to the map-making problem is obtained by maximizing the likelihood of the data given a noise model \citep{wright96,tegmark97,stompor02}. The solution, resulting in a minimum variance map, is given by the well-known generalized least squares (GLS) equations, which provide an estimate of the sky signal and its covariance
\begin{eqnarray}
	\hat{\mathbf{T}} 
	& = &
	\left(\mathbf{A}^T \mathbf{N}^{-1}\mathbf{A}\right)^{-1} \cdot \mathbf{A}^T \mathbf{N}^{-1}\mathbf{d} 
	\label{eq:mapmaking}
	\\
	{\cal N} 
	& = &
	\left(\mathbf{A}^T \cdot\mathbf{N}^{-1}\cdot \mathbf{A}\right)^{-1} \ .
	\label{eq:mapmaking_covmat}
\end{eqnarray}

In practice, the inversion or even the calculation of the $N_p \times N_p$ matrix $\mathbf{A}^T\mathbf{N}^{-1}\mathbf{A}$ is impossible for large datasets such as Planck, where the number of sky pixels, $N_p$, can be as large as many millions. Thus, Eq.~\ref{eq:mapmaking} is usually solved using iterative methods such as the preconditioned conjugate gradient (PCG) \citep{golub96} and capitalizes on fast Fourier transforms (FFT) to perform the Toeplitz matrix-vector products~\citep{doré01, degasperis05, cantalupo10}.

\subsection{Destriping methods}

So-called ``destripers'' attempt to simplify the general map-making problem described above for Planck-like scanning strategy when the solution would require too many resources. The destriping technique for the CMB map-making has been investigated in minute detail for the Planck satellite \citep{burigana99, delabrouille98, maino99, revenu00, maino02, keihanen04, stompor04}. This technique takes advantage of the detectors observing large circles on the sky when the Planck telescope is spinning. Each circle is observed several times (of order 50) consecutively. Averaging over these circles enables us to compile datasets of higher signal-to-noise ratio. It has been shown by \citet{janssen96} that the low frequency noise can be represented by a uniform offset on a given baseline, corresponding in the case of Planck to a stable pointing period called a ring. 
However, it can also be adapted to ground-based or balloon-borne experiments \citep{sutton10}. 

In the destriping approach, the noise is approximated as a low-frequency component represented by the offsets $\mathbf{x}$ unfolded to time-ordered data by the offset `pointing' matrix $\mathbf{\Gamma}$ and remaining Gaussian noise $\mathbf{n}$ characterized by a correlation matrix, ${\mathbf N}$
\begin{equation}
	\mathbf{d} 
	= 
	\mathbf{A} \cdot \mathbf{T} + \mathbf{\Gamma}\cdot\mathbf{x} + \mathbf{n}.
\label{eq:destDataModel}
\end{equation}
The matrix $\mathbf{\Gamma}$ assigns to every consecutive piece of the time stream data the respective offset amplitude and its elements are therefore either $1$ or $0$.
The maximum-likelihood estimates of the offset amplitudes, $\mathbf{x}$, can be found from the time-ordered data, $\mathbf{d}$, by solving
\begin{equation}
	\left( \mathbf{\Gamma}^T \mathbf{N}^{-1} \mathbf{Z} \mathbf{\Gamma} \right) 
	\cdot \mathbf{x} 
	= 
	\mathbf{\Gamma}^T \mathbf{N}^{-1} \mathbf{Z} \cdot \mathbf{d},
\label{eq:destriping}
\end{equation}
where
\begin{equation}
	\mathbf{Z} 
	= 
	\mathbf{I} 
	- 
	\mathbf{A} \left(\mathbf{A}^T \mathbf{N}^{-1} \mathbf{A}\right)^{-1} 
	\cdot \mathbf{A}^T \mathbf{N}^{-1}
\end{equation}
and the covariance matrix for offsets reads
\begin{equation}
	\mathcal{C}_{\mathbf x}
	= 
	\left( \mathbf{\Gamma}^T \mathbf{N}^{-1} \mathbf{Z} \mathbf{\Gamma} \right)^{-1}.
\label{eq:offCov}
\end{equation}
We note that, unlike Eq.~\ref{eq:destriping}, which provides an unbiased, though potentially non-optimal, estimate of the offset amplitudes even if the assumed and actual properties of the instrumental noise differ, this last equation holds only if our data model given by Eq.~\ref{eq:destDataModel} and the assumed noise correlation, ${\mathbf N}$, are both correct. This is worth emphasizing as in the destriper applications one usually assumes that the remaining noise, ${\mathbf n}$, is uncorrelated and typically piece-wise white, and thus the matrix ${\mathbf N}$ is diagonal. This is indeed a key assumption from which the performance advantage of the destriper technique stems~\citep[][]{ashdown09}. Once the offsets, ${\mathbf x}$, are estimated, we can calculate the sky map, by coadding samples into pixels after first subtracting the estimated offsets from the TODs, i.e.,
\begin{equation}
{\mathbf m} = \left({\mathbf A}^t \, {\mathbf N}^{-1} \, {\mathbf A}\right)^{-1} \, {\mathbf A^t} \, {\mathbf N}^{-1} \, \left({\mathbf d} \, - \, {\mathbf \Gamma} \, {\mathbf x}\right).
\label{eq:destMap}
\end{equation}
Again such an estimate of the sky signal will be unbiased even if the simplifying assumptions about the noise, ${\mathbf n}$, are adopted. If these are correct, or sufficiently accurate, then the error covariance matrix of the destriper map can be calculated as
\begin{equation}
{\cal C}_{\bf m} \,  = \, \left({\bf A}^t \, {\bf N}^{-1} \, {\bf A}\right)^{-1} \, + \, \left({\bf A}^t \, {\bf N}^{-1} \, {\bf A}\right)^{-1} \, {\bf A}^t \, {\bf N}^{-1} \, {\bf \Gamma} \;
\left( \mathbf{\Gamma}^T \mathbf{N}^{-1} \mathbf{Z} \mathbf{\Gamma} \right)^{-1}
 {\bf \Gamma}^t \, {\bf N}^{-1} \, {\bf A}\, \left({\bf A}^t \, {\bf N}^{-1} \, {\bf A}\right)^{-1}.
\label{eq:destMapError}
\end{equation}
Equations~\ref{eq:offCov} \&~\ref{eq:destMapError} can clearly provide at least some insights into the error correlation structure of the recovered maps and even be suitable for at least some of their statistical analyses, even if the reminder noise, ${\mathbf n}$, is not strictly speaking uncorrelated and for this reason their calculation is included in the software described here and discussed later on.

As shown in \citet{poutanen06} and \citet{ashdown07b}, the baseline length used to define the offsets need not be tied to the length of a Planck ring. Including priors on the low frequency noise, the destriping algorithm is equivalent to the GLS algorithm in the short baseline limit. In the Planck case, the duration of the stable pointing period is not constant over the time so that the baseline lengths vary slightly from 40 to 65 min. However, as we show later, this generalization of the \citet{janssen96} prescription does not affect the destriper performances.

\section{Photometric calibration}
\label{sec:calibration}

The bolometer signal measured through I-biasing is proportional to the small variation in the incoming power from the sky. To express the measurement in sky temperature units, one has to determine a gain per detector based on a known source in the sky. 
For low frequency channels (roughly between $20$ and $300$~GHz) for which the CMB signal is sufficiently high, the CMB dipole is usually used as a primary calibrator for experiments with large sky coverage. This is because it is only marginally affected by pointing errors and beam uncertainties, is a stronger signal than CMB anisotropies (by a factor $100$) but not bright enough to cause non-linearities in the detectors, and has the same electromagnetic spectrum.
At higher frequency (typically above $300$~GHz), calibration of data on Galactic signal, based on ancillary data \citep[e.g. FIRAS maps][]{firas}, is preferred. For smaller sky coverage, calibration is usually derived from point source objects of known flux.

The CMB dipole is induced by the Doppler effect of the relative motion of the satellite with respect to the last scattering surface. The solar system motion with respect to the last scattering surface (referred to as the {\it solar dipole}) is the dominant component of the satellite velocity. A residual contribution (called hereafter the {\it orbital dipole}) is induced by the motion of the satellite with respect to the solar system. The solar dipole can be considered as sky stationary during the observations and is thus projected onto the sky as an $\ell=1$ component with an amplitude measured by WMAP of $3.355 \pm 0.008$~mK \citep{wmap05}. Relativistic corrections to the solar dipole produce second order anisotropies at higher multipoles with amplitudes proportional to $\beta^\ell$ and more importantly couple both the dipole components as we discuss later on. Though the orbital dipole flux is typically one order of magnitude lower than the solar dipole, it is time dependent, and its time-variability is precisely determined by the satellite velocity. 

The orbital dipole signal is modulated on a one year period. As it is not projected onto the sky, the ring-ordered data, $\mathbf{d}$, for each sample $s$ is modeled as
\begin{equation}
\mathbf{d} 
= 
\mathbf{A} \cdot \mathbf{T} + g \times \mathbf{t}_{orbital} + \mathbf{n},
\end{equation}
where $\mathbf{A}$ is the pointing matrix relating ring pixels to those of the sky, and $\mathbf{t}_{orbital}$ is the time-dependent orbital dipole signal. The unknown are the sky temperature $\mathbf{T}$ (including the solar dipole) and the gain $g$.

As $\mathbf{t}_{orbital}$ is known, the calibration problem is linear and can be solved directly in the same way as the destriping problem. For instance, the maximum-likelihood estimate of the coefficient $g$ can be obtained by marginalizing for $\mathbf{T}$ over the sky solving the equation
\begin{equation}
\left( \mathbf{t}_{orbital}^T \mathbf{N}^{-1} \mathbf{Z} \mathbf{t}_{orbital} \right) \times g
= 
\mathbf{t}_{orbital}^T \mathbf{N}^{-1} \mathbf{Z} \cdot \mathbf{d},
\label{eq:calibration}
\end{equation}
where
\begin{equation}
\mathbf{Z} 
= 
\mathbf{I} 
- 
\mathbf{A} \left(\mathbf{A}^T \mathbf{N}^{-1} \mathbf{A}\right)^{-1} 
\cdot \mathbf{A}^T \mathbf{N}^{-1},
\end{equation}
and subsequently used to estimate the sky map.

For Planck, using polarization-sensitive detectors to solve such a calibration problem for a single detector at a time turns out to be insufficient, as the map-making problem is either degenerate if polarization is explicitly modeled, or the answer biased, if polarization is neglected (see Sect.~\ref{sec:calib_syste}). However including multiple detectors in the calibration problem in Eq.~\ref{eq:calibration} leads to a set of non-linear equations. To avoid this problem, our algorithm performs only a single detector gain estimation at any given time, although it does this for a few detectors in parallel, the calibrated data of which are then used to determine offset estimates. The latter are subsequently used for the calibration estimates resulting in an iterative procedure.

\section{Implementation}
\label{sec:polkapix}

In the context of Planck data analysis, several destriper codes have been developed, such as MADAM \citep{keihanen05,keihanen10} and Springtide \citep{ashdown07a}. In this paper, we present a modular code called {\sc Polkapix} based on four separate steps: the ring-making, the offset determination, the photometric calibration, and the projection of the rings onto a map. The offset determination and the calibration are solved iteratively. We note that explicit ring-making is not necessary in destriper codes, although we use it not only as a convenient means of compressing the data but also as intermediate stage products that are useful for monitoring the systematics. In addition, and for the same reason, our implementation allows for internal, thus potentially different, sky resolutions for the offset and calibration determination independently of the one used for the final map projection.

\subsection{Ring-making}

Rings are partial sky maps produced via a projection onto the sky of each single pointing period separately. These rings therefore provide a compressed and higher signal-to-noise ratio rendition of the original TODs. The length of each stable pointing period varies between 40 and 65 min (with a mean of 45 min) in the Planck scanning strategy. During each period, noise is assumed to be white, with the low frequency part folded in as an overall ring offset, and consequently the rings are computed using a simple noise-weighted projection procedure. The resulting noise in the ring domain is thus (nearly) white, modulo the offset, and mostly uncorrelated from one ring to another.

For calibration purposes, as in Eq.~\ref{eq:calibration}, the orbital dipole has to be averaged in the same way for each pixel of the ring. This is done using the pointing direction and the satellite velocity calculated at each sample. As the dispersion of the observed direction falling into each ring pixel is very low (below $10^{-3}$), the averaging effects of pixelization are found to be negligible and no loss of accuracy is incurred as a result.

To avoid introducing any additional binning of the data, we choose a sky pixelization as a basis for this ring making \citep[HEALPix,][]{healpix}. Our ring structure is therefore called HPR for HEALPix Rings in the following. As the ring are produced via simple projection assuming white noise only, samples that are flagged as bad, are simply not included in the ring making procedure, without any need for additional processing, such as gap-filling~\citep[e.g.,][]{stompor02}.

\subsection{Offsets determination}
\label{sec:polkapixOff}

The destriping algorithm solves Eq.~\ref{eq:destriping} for the offsets $x$ (one offset per ring) taking as an input a set of HPR, $\mathbf{d}$, computed at the ring-making stage.
We also impose the external constraint $\sum \mathbf{x}=\hbox{\sc const}$ to break the degeneracy in the offset determination and set the arbitrary constant to $0$ so that the mean value of the TODs (unphysical in any case) is conserved. 
Though we typically use the same underlying pixelization, HEALPix~\citep[][]{healpix}, for the ring and sky signals, these may have different resolutions for the intensity and polarization parts. The resolution should be high enough for the sky signal to be considered constant across the pixel, thus typically lower than the characteristic instrumental beam scale. However, a trade-off should be made between the offsets uncertainty related to the level of noise in each pixel (which decreases for larger pixels) and the correlation between offsets (demanding smaller pixels). Moreover, to satisfy the first criterion above it is typically necessary to mask the inner part of the Galaxy, while estimating the offsets, as the galactic gradients are often too strong, leading to a poor estimate of the sky signal. These questions are discussed in more detail in Sect.~\ref{sec:destriping_results}.

Two solvers of Eq.~\ref{eq:destriping} have been implemented here: the full inversion approach, allowing for, and requiring, the estimation of the offset covariance matrix $\mathcal{C}$ and a (much faster) iterative method through conjugate gradient.
The code has been implemented within the HFI Data Processing Center infrastructure, and is fully parallelized as far as both the workload and memory consumption are concerned. It is able to deal simultaneously with data from several detectors at the same frequency, as needed, for instance, to reconstruct polarization sky. As an example, the destriping is performed in a few minutes for 4 detectors on 32 processors using the full simulation described in Sect.~\ref{sec:simulations}.
The reconstruction of the polarization for each pixel requires the inversion of each three-by-three elements of the $(\mathbf{A}^T \mathbf{N}^{-1}\mathbf{A})$ matrix, some of which may be ill-conditioned for pixels with a insufficient number of bolometer orientations \cite[see][]{ashdown07a}. We considered only pixels for which the condition number for this three-by-three matrix is lower than $10^3$.

The code described in this work was used to produce the Planck Early Release \citep{edp} and is one of the workhorses of the Planck-HFI Data Processing Centers. The implemented algorithm is overall very close to that of Springtide as discussed in detail in \citet{ashdown07a}. We checked that indeed both these implementations produce the same offsets on a simple simulation of 4 HFI detectors at 143~GHz and including only the CMB signal (see Sect.~\ref{sec:simulations} for details on simulations). The residual maps, i.e., a difference between noiseless input and output maps, give very similar r.m.s. with relative differences being: $\sim 10^{-6}$ for $I$ and $\sim 5\cdot10^{-5}$ for $Q$ and $U$. The r.m.s. offsets differences agree within a relative error of $\sim 5\cdot10^{-8}$.
For a detailed comparison of map-making methods, we refer the reader to the papers of the Planck-CTP Working Group \citep{ashdown07a,ashdown07b,ashdown09}.

\subsection{Dipole calibration}

The calibration module solves Eq.~\ref{eq:calibration} for gain $g$ for each of the included bolometers individually. As an input, it takes pre-computed rings corrected for the offset estimates, $x$, as derived earlier. The code is parallelized so that the gain estimation is achieved in fewer then two minutes on $16$ processors. We use an underlying sky pixelization for intensity based on HEALPix and set its resolution as discussed in Sect~\ref{sec:calibration_results}. The calibration error in the fitted gain is then estimated as 
\begin{equation}
	\sigma_g^2 = \mathcal{F}^{-1}_{cal} = \mathbf{t}_{orbital}^T \mathbf{N}^{-1} \mathbf{Z} \mathbf{t}_{orbital}.
\label{eq:fisher_cal}
\end{equation}
We note that this equation is an approximation neglecting the errors due to an offset determination. However, as we use multiple detectors for the latter, the approximation is expected to be very good.

The calibration algorithm takes advantage of the orbital dipole not being fixed on the sky, unlike the solar dipole. In practice, relativistic corrections couple solar and orbital dipoles creating a additional non-stationary signal. Though the latter is three orders of magnitude ($\beta=10^{-3}$) below the orbital dipole signal, we have to include the relativistic correction in the calibrator. We measure a relative bias of $6\times10^{-6}$ to the recovered gain when using the orbital dipole, instead of the non-stationary signal (with coupling corrections).
Given the statistical error ($\sim 5 \times 10^{-5}$, see Sect.~\ref{sec:calib_stat}), we conclude that the orbital dipole can be safely used as a calibrator for Planck-HFI data even if, for the rest of the study, we use the exact non-stationary calibrator constructed as the difference between the CMB dipole (including solar, orbital, and relativistic corrections) and the solar dipole.

The procedure may be susceptible to pixelization effects caused by signals with large intra-pixel variations such as those found in the Galactic plane or for point sources. We evaluate the impact of this effect using a Galactic mask in Sect.~\ref{sec:calib_syste}.
The calibration code can also provide a constraint on the gain due to the data corresponding to each sky pixel separately, as well as an estimate of its uncertainty. The overall calibration constraint is then a weighted average of the these pixel-specific values, with the weight being its estimated uncertainty. Thanks to this facility, we can identify the sky areas that contribute most significantly to the final result of the calibration procedure.

\subsection{Rings projection}

Once the offsets have been estimated, they are subtracted from each ring, which removes the low frequency noise component from the data. A ``corrected" map is then obtained by simply coadding ring-pixel amplitudes corresponding to the same sky pixels and weighted by the estimated ring-pixel white noise estimate. We note that though the resolution of the final map does not have to be the same as that of the rings it should be typically no higher than that. In this paper, we always project the final full-sky maps on high resolution (nside = 1024), whereas we study the impact of degrading the resolution on the underlying maps in both destriping and calibration.

\section{Simulations}
\label{sec:simulations}

Time-ordered data (TOD) were simulated for $380$ days of observations corresponding to slightly more than two sky surveys. The sampling frequency for HFI was set to $180$~Hz, so that the total number of samples is about $5.6 \times 10^9$ per bolometer. For $12$ detectors and the addition of the pointing information, the total volume of the data amounts to $700$~Gb per simulation. We restrict our discussion to four bolometers only per channel, which allows for a polarization reconstruction while limiting the overall data size to one more readily manageable. Nevertheless, we rescale the noise level so that noise in the final map is comparable to that expected for a complete Planck-like channel map. We focus on two CMB channels centered at $143$ and $217$~GHz.
Simulations were done using the Level-S simulation codes \citep{reinecke06} ported into the DPC-HFI infrastructure, and in particular the Planck I/O library. Beam effects were simulated assuming Gaussian circular beams.

\subsection{Scanning strategy}
\begin{figure}[h!]
	\center
	\includegraphics[width=0.49\textwidth]{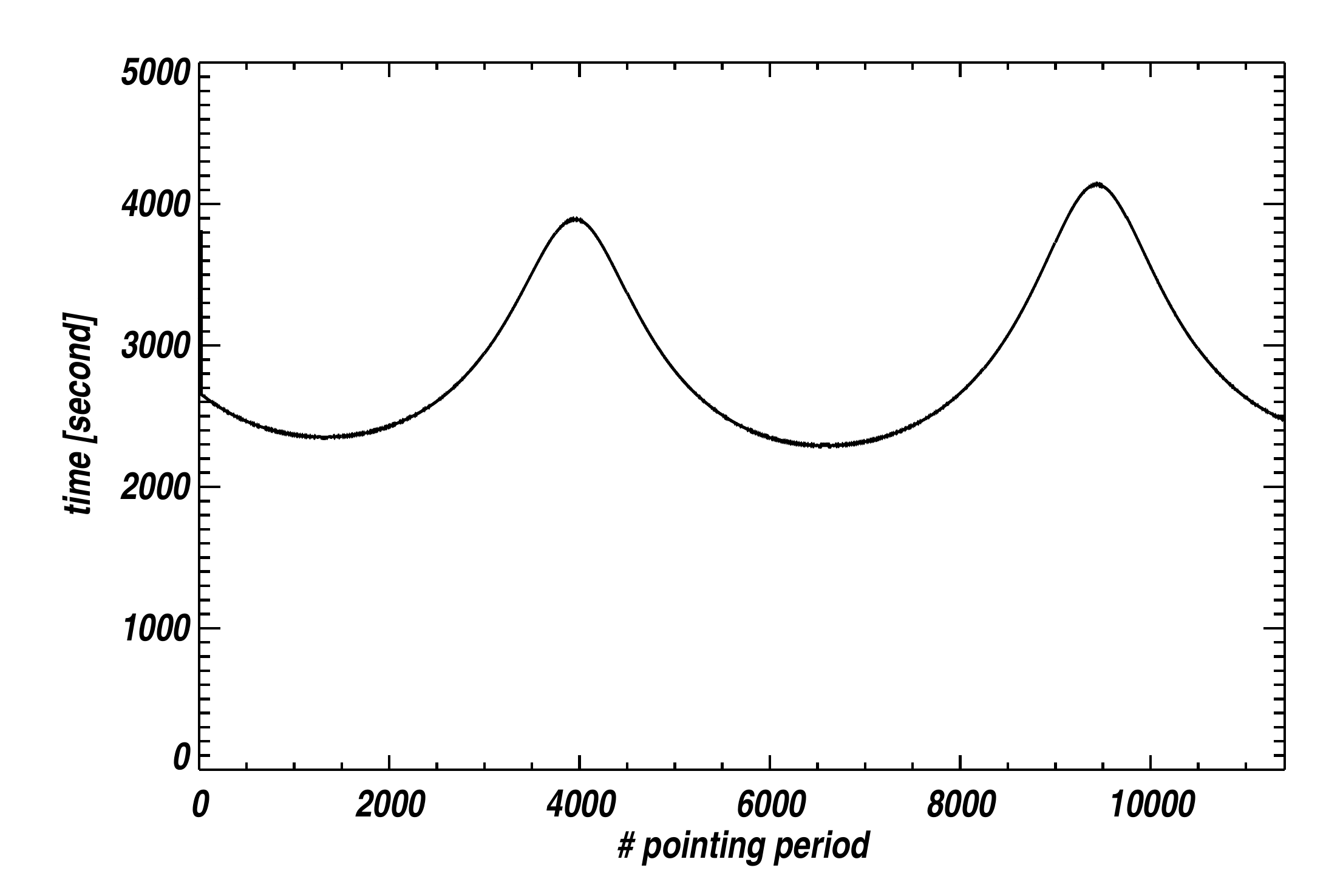}
	\caption{Dwelling time of the simulated stable pointing periods for the $12.5$ month-long observation. The variation in the period length reflects that the spin axis does not remain in the Ecliptic plane but follows a cycloid path.}
\label{fig:pointing_periods}
\end{figure}

\begin{figure}[h!]
	\center
	\includegraphics[width=0.49\textwidth]{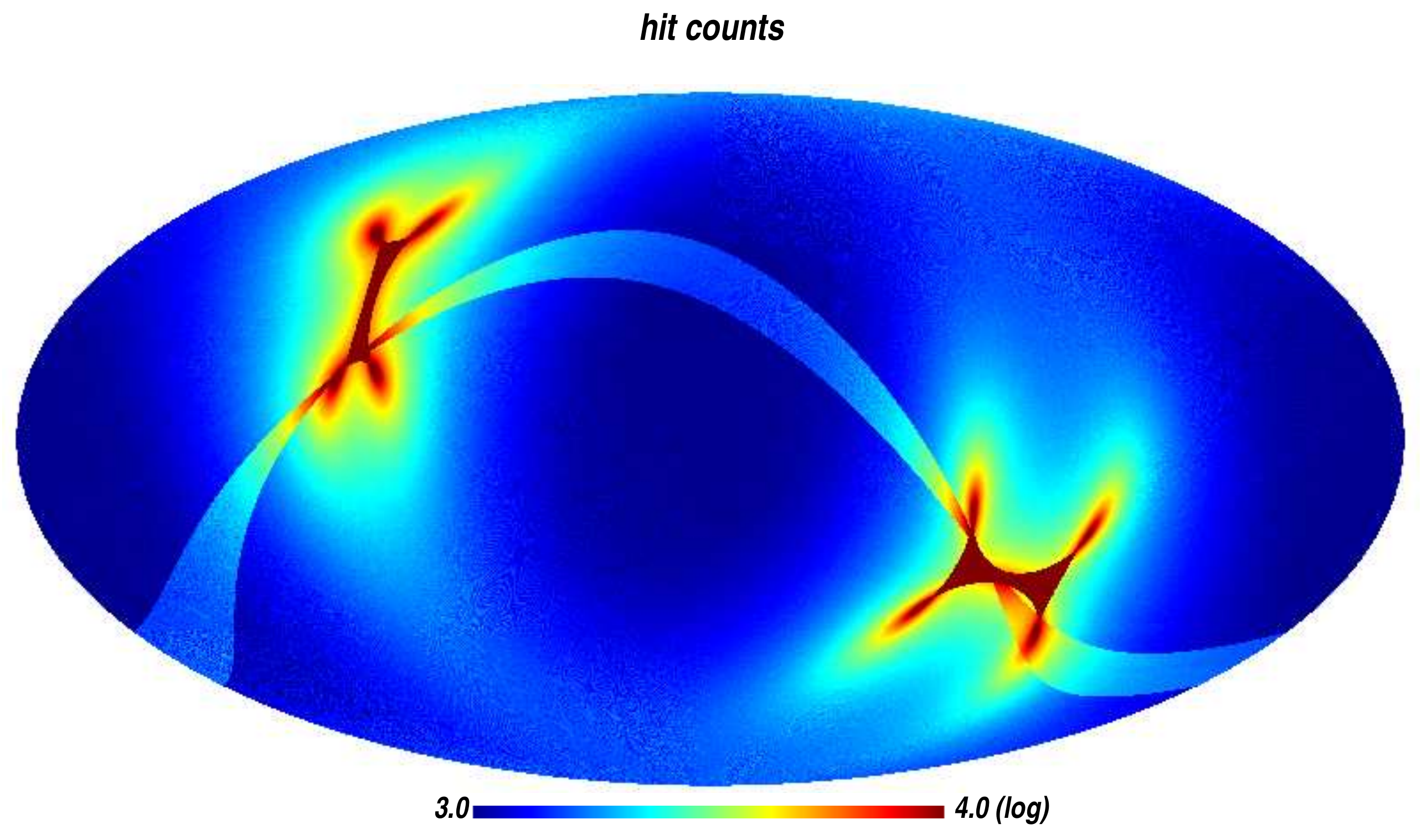}
	\caption{A map of hit counts per $3.4$~arcmin pixels (Nside $=1024$) in Galactic coordinates corresponding to $12.5$ months of observation for four detectors. The sky is covered at least twice. Redundancy of the revisits is maximal around the Ecliptic poles. The conspicuous s-shaped band with a higher hit count corresponds to the beginning of the third survey.}
\label{fig:hitcounts}
\end{figure}

We used a Planck-like cycloidal scanning strategy. The satellite was assumed to be spin-stabilized and rotates on its axis once per minute. 
The spin axis follows a cycloidal path on the sky by step-wise displacements of approximately 2~arcmin every $48$ min. The stable pointing period between two repointings is not constant in order to take into account the spin axis not remaining in the Ecliptic plane but following a cycloid path (Fig.~\ref{fig:pointing_periods}). Between the repointings, the spin axis nutates with a mean amplitude of $2$~arcsec. Detectors scan the sky following almost great circles as they point at $83.8$\deg away from the spin axis (Fig.~\ref{fig:hitcounts}). We also include small variations in the spin rate (rms $2.16$arcsec/s).

For the polarization studies, we simulated four bolometers at $143$~GHz coming from two horns of the Planck-HFI focal plane (143-1 and 143-3). Each horn groups two detectors (labeled $a$ and $b$) sharing the same pointing with a polarized grid orientation rotated by $90$\deg. The two horns follow the same path on the sky but are $2$\deg~apart from each other. Their relative orientation of the polarized grids in each horn differ by $45$\deg. During the $12.5$ months of the mission, the four detectors observe each pixel on the sky with multiple polarization orientation in a $3.44$~arcmin resolution map (HEALPix nside$ = 1024$). Therefore we are able to determine the three Stokes components in each pixel. In addition, we simulated one total power, $217$~GHz bolometer.

\subsection{Sky signals}

The simulated signal combines the CMB and diffuse Galactic foregrounds from the Planck Sky Model\footnote{See http://www.planck.fr/heading79.html} at a resolution of 1.7~arcmin (HEALPix nside = 2048). The CMB anisotropies are generated using CMBFAST~\citep{cmbfast} with the WMAP five-year best-fit model without lensing ($\Omega_m=0.257,\Omega_b=0.044,\Omega_\Lambda=0.743,h=0.719,\tau=0.1,n_s=0.963,r=0.02$). Both solar and orbital dipoles are included. The Galactic foreground signals include thermal and spinning dust, synchrotron radiation, and free-free scattering in intensity. Polarized signals for thermal dust and synchrotron are also added. More details about the sky model used for this work can be found in \citep{leach08} and \citep{betoule09}.

The maps were smoothed using a symmetric Gaussian beam with FWHM of $7.05$ and $4.72$~arcmin at $143$ and $217$~GHz, respectively. TODs were then generated using Level-S codes \citep{reinecke06} extracting the signal from the input sky maps according to the scan strategy. No bolometer time constant was included.

\subsection{Noise properties}

Streams of $1/f^\alpha$-noise were generated at the TOD level using the algorithm described in \citet{plaszczynski07}. We selected one of two values of the slope ($\alpha= 1$ and $2$) and two knee frequencies ($f_{knee}= 0.1$ and $0.01$~Hz), with a cutoff frequency $f_\text{min}=10^{-5}$Hz below which noise becomes white. In this approach, a white Gaussian noise realization is generated and the long range correlations are obtained through a linear digital filter. For a given \textit{seed} of the generator, one can directly study the effect of the pure $1/f^\alpha$-noise component using the difference of reconstructed maps with and without the correlation turned on.
For Monte Carlo simulations, we need a significant number of noise realizations, which renders impractical the writing of the TOD to disk. We therefore directly project the noise at HPR level using a pre-computed HPR pointing matrix.

The thermal fluctuation expected for Planck-HFI detectors are usually modeled by a $1/f^2$ Fourier spectrum with a low knee frequency. The noise power spectra of HFI bolometers are more accurately described by a $1/f$ spectrum ($\alpha=1$) with $f_{knee}\simeq0.1$~Hz plus white noise \citep{edp}. We set the TOD white noise level in order to match the expected noise level of the foreseen {\it Planck}-HFI, $143$~GHz frequency map. Thus, the noise r.m.s. per sample was set to $589\ \mu K$ in thermodynamic CMB scale \citep{bluebook}.
The noise covariance matrix in the map domain is given by Eq.~\ref{eq:mapmaking_covmat}. If correlated errors are small, the covariance matrix for ($I$, $Q$, $U$) at pixel $p$ reads
\begin{equation}
	{\mathbf M}_p = 
	\frac{1}{\sigma^2}
	\left( 
		\begin{array}{ccc} 
		\sum{1} & \varepsilon \sum {\cos(2 \alpha_i)} & \varepsilon\sum {\sin(2 \alpha_i)} \\ 
		\varepsilon \sum {\cos(2 \alpha_i)} & \varepsilon^2 \sum {\cos^2(2 \alpha_i)} & \varepsilon^2 \sum {\cos(2 \alpha_i)\sin(2 \alpha_i)} \\ 
		\varepsilon \sum {\sin(2 \alpha_i)} & \varepsilon^2 \sum {\cos(2 \alpha_i)\sin(2 \alpha_i)} & \varepsilon^2 \sum {\sin^2(2 \alpha_i) }
	\end{array} \right )^{-1}
	=
	\sigma^2
	\left(
		\begin{array}{ccc} 
		\frac{1}{n_p} & 0 & 0\\
		0 & \frac{2}{\varepsilon^2\, n_p} & 0 \\ 
		0 & 0 & \frac{2}{\varepsilon^2\, n_p}
	\end{array} \right )
	\label{eq:pixNoise}
\end{equation}
where the sums extend over all samples of all considered detectors falling into the pixel $p$, and we have assumed that the polarization efficiency, Eq.~\ref{eq:polDataModel}, is the same for all detectors. The second step results from the cancellation of the off-diagonal contributions coming from two detectors of each horn, which thus have their polarizers rotated by $90$ degrees with respect to each other. As the polarization efficiencies are not all the same and the polarizers are not rotated by exactly $90$ degrees, the second step is only an approximation with the off-diagonal elements being on the order of ${\cal O}(\Delta \epsilon)$ and ${\cal O}(\Delta \alpha)$, where $\Delta \epsilon$ denotes a difference between the polarization efficiencies of two detectors of the same horn and $\Delta \alpha$ the deviation from orthogonality. The off-diagonal terms are further suppressed for sky pixels, which are observed by the same horn multiple times with different attack angles (the correlation is lower than $5$\% for our scanning strategy). Consequently, the matrices ${\mathbf M}_p$ are usually diagonally dominated for sufficiently well-observed pixels, with the diagonal elements being approximately given by Eq.~\ref{eq:pixNoise}.
Given the hit counts per pixel $n_p$ (Fig.~\ref{fig:hitcounts}) and the characteristics of the HFI detectors that we used ($\varepsilon=$0.83, 0.85, 0.84, 0.90), the averaged white noise per pixel we expect in our simulations is therefore 15.4~$\mu K_{CMB}$ for $I$ and 25.6~$\mu K_{CMB}$ for $Q$ and $U$ maps at 143~GHz. In the following, we compare the level of residuals with those values.

\section{Destriping parameters and performances}
\label{sec:destriping_results}

\subsection{Systematic studies}
\label{sec:destriping_syste}

We now investigate the various systematic effects possibly affecting the offset determination. We use a simulated sky with CMB and Galactic signal at $143$~GHz for four bolometers, allowing for intensity and polarization reconstruction. For each bolometer, we generate 100 pure white noise realizations. HPR are constructed at high resolution (nside = 2048) and the final full-sky map is projected at nside = 1024. We first built the signal map using coaddition in each pixel. We also compiled the white noise map for each realization without destriping. For each noise realization, we constructed the residual map subtracting the signal+white noise map from the destriped map. We then computed full-sky angular power spectra for each residual map using the {\it Xpol} \citep{xpol} code, based on the S$^2$HAT library\footnote{http://www.apc.univ-paris7.fr/$\sim$radek/S2HAT.html}~\citep[][]{s2hat, szydlarski11} and averaged them over the simulated realizations.

\begin{figure*}[h!]
	\center
	\includegraphics[width=\textwidth]{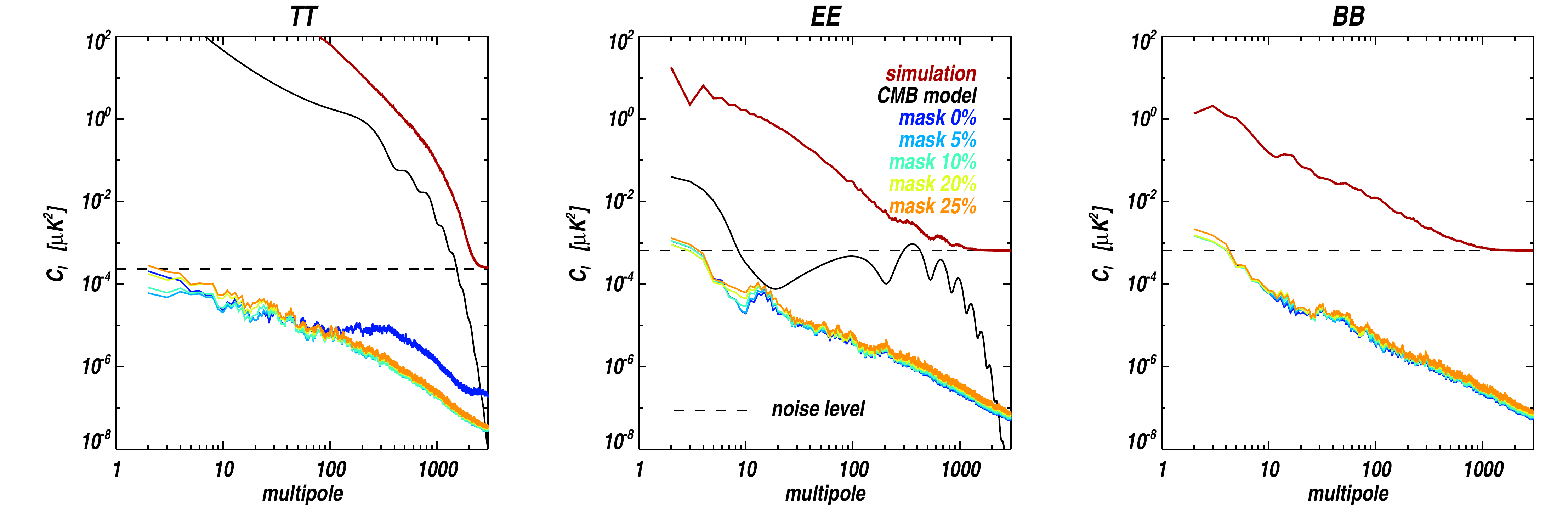}
	\caption{Noise residual full-sky power spectra with various internal Galactic mask applied to the offset determination. Residual maps are at nside = 1024. Spectra are smoothed over five multipoles. Signal spectra are plotted in {\it red}. As a reference, we also plot the CMB WMAP five-year fiducial model ({\it black}) and a total power of the simulated sky signal, CMB+Galaxy+noise, ({\it red curve}). {\it From left to right:} temperature, E-mode, and B-mode.} 
\label{fig:mask_pol}
\end{figure*}

First we investigate the effect of large signal variations within the pixels on the offset determination. These pixels can be found mostly in the Galactic plane so we remove the pixels with the highest Galactic signal using a mask (based on FIRAS intensity maps) in the offset determination. We gradually increase the masked-out sky fraction from $0$\% to $25$\%. We use the destriping module to estimate the offsets using an internal sky resolution of 6.9~arcmin (HEALPix nside = 512). The results are shown in Fig.~\ref{fig:mask_pol}. We can see that only temperature residuals are sensitive to the masking and that a $5$\% mask is sufficient to suppress pixel effects due to strong sub-pixel signal gradients caused by the Galactic signals. Consequently for the following, we use an internal $5$\% Galactic mask at $143$~GHz for offset determination.

\begin{figure*}[h!]
	\center
	\includegraphics[width=\textwidth]{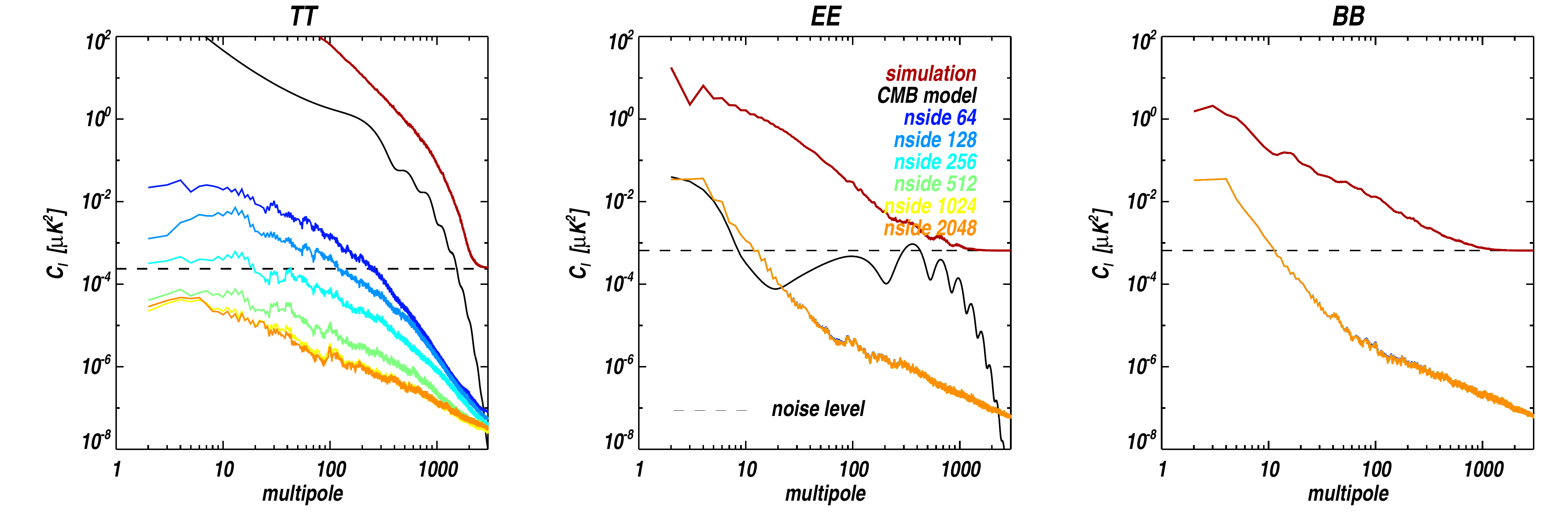}
	\caption{Noise residual full-sky power spectra with various internal resolutions (nside = 64, 128, 256, 512, 1024) and neglecting polarization in the offset determination. Residual maps are at nside = 1024. Spectra are smoothed over five multipoles. Signal spectra are plotted in {\it red}. As a reference, we also plot the CMB WMAP-5yr fiducial model ({\it black}). The horizontal, dashed lines show the predicted noise level. {\it From left to right:} temperature, E-mode, and B-mode.} 
\label{fig:residuals_nside}
\end{figure*}

Subsequently, we assess the impact of the assumed sky resolution on the offset determination procedure. We do this in two steps. First, we neglect the polarization signal while determining the offsets, and consider several internal resolutions from nside = $64$ to $2048$ in the offset determination. Figure~\ref{fig:residuals_nside} shows the power spectra of the destriping residuals for a pure white noise. In terms of temperature, at low internal resolutions (nside lower than $256$), residuals due to the pixelization effect dominate at low multipoles over the instrumental noise level. Residual power spectra in polarization are unaffected by changes in the internal resolution for the temperature and are consistently higher at low-$\ell$s than the noise level. We attribute this last observation to no polarization being included during the offset computation step.

\begin{figure*}[!h]
	\center
	\includegraphics[width=\textwidth]{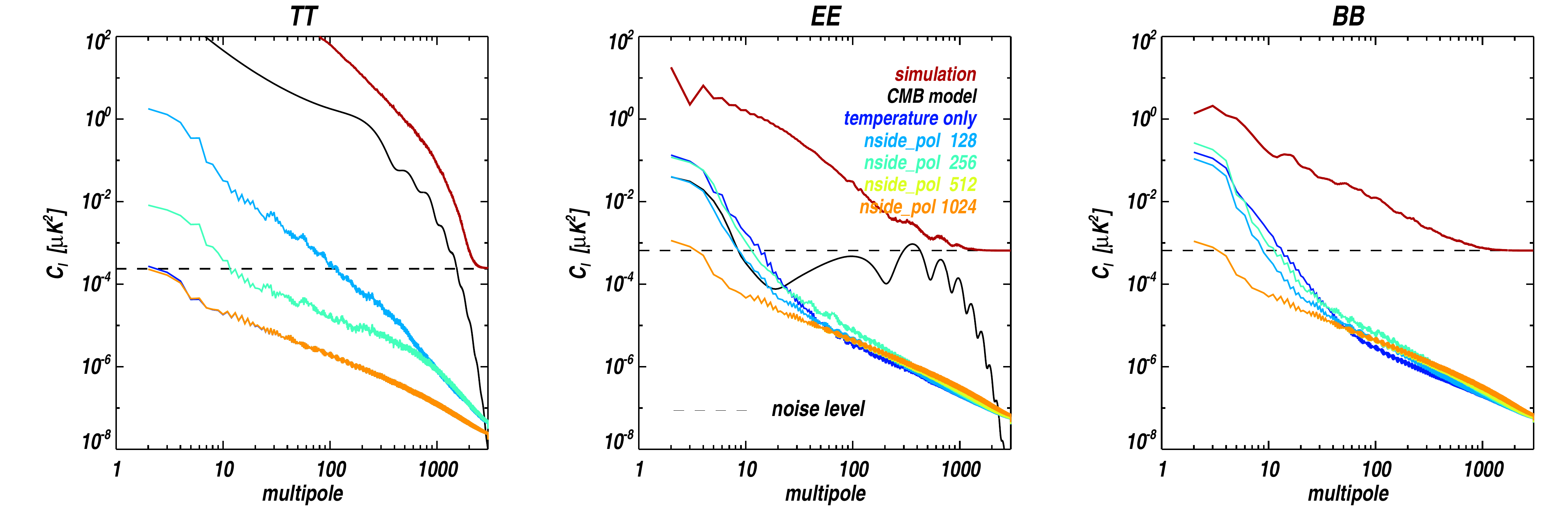}
	\caption{Noise residual full-sky power spectra when using different internal sky resolution for polarization in the offset determination. Internal resolution for intensity is fixed to nside = 1024. Residual maps are at nside = 1024. Spectra are smoothed over five multipoles. Signal spectra are plotted in {\it red}. As a reference, we also plot the CMB WMAP five-year fiducial model ({\it black}). {\it From left to right:} temperature, E-mode, and B-mode.}
	\label{fig:polar_destriping}
\end{figure*}

Finally, we compare the destriping performances when the polarization is included (for various internal sky resolutions) in the offset determination. For this comparison, the internal resolution for intensity is fixed at $3.4$~arcmin (HEALPix nside = $1024$). As shown in Fig.~\ref{fig:polar_destriping}, including the polarization reduces the level of the residuals of the $E$ and $B$ mode polarized spectra. For polarization, at high resolution (nside greater than $512$), we found residuals no larger than the white noise level over the entire $\ell$ range, including the low multipole range ($\ell < 20$). At higher multipoles, a higher resolution seems to produce a slight increase in the residuals but we understand this is negligible given the white noise level. As far as the intensity is concerned, a low resolution for polarization induces strong residuals in the intensity map for multipoles lower than $\ell=100$.

From these studies, we conclude that polarization must be taken into account during the destriping, even if this requires more data (for Planck at least four bolometers) leading to an increase in the CPU time as well as the introduction of possible systematics due to detector mismatches (not discussed here). Moreover, we have demonstrated that a resolution of 6.9~arcmin (HEALPix nside = 512) for both intensity and polarization ensures that the residuals remain negligible relative to white noise level.

\subsection{Statistical uncertainties and offset covariance matrix}
\label{sec:covmat}

\def\nsimu{10000}

The statistical uncertainties in the offset determination are quantified by the offset-covariance matrix, $\mathcal{C}_{\mathbf x}$, which in general depends  not only  on the scanning strategy but also instrumental characteristics, both of which enter via the pointing matrices $\mathbf{A}$ and $\mathbf{\Gamma}$, Eq.~\ref{eq:offCov}, as well as noise properties of the detectors encoded in the matrix $\mathbf{N}$. However, this is typically the scan pattern, which plays a dominant role in determining the overall  structure of the noise covariance \citep[e.g.][]{stompor04, efstathiou07}, and conversely most of the major features of the covariance can be traced back to the details of the scanning. Below we discuss the results of the numerical computations of $\mathcal{C}_{\mathbf x}$, examples of which are displayed in Figs.~\ref{fig:covmat_nside}, \ref{fig:covmat_mask}, \ref{fig:covmat}, and~\ref{fig:covmat_multibolo}, emphasizing this connection. To this end, we first rewrite Eq.~\ref{eq:offCov} more explicitly as
\begin{equation}
	{\cal C}_{\bf x} = 
		\left({\bf \Gamma}^t {\bf N}^{-1} {\bf \Gamma}\right)^{-1} + 
		\left({\bf \Gamma}^t {\bf N}^{-1} {\bf \Gamma}\right)^{-1} \left( {\bf \Gamma}^t {\bf N}^{-1} {\bf A} \right) {\cal C}_{\bf m} \left( {\bf A}^t {\bf N}^{-1} {\bf \Gamma} \right) \left({\bf \Gamma}^t {\bf N}^{-1} {\bf \Gamma}\right)^{-1},
\label{eq:offCovLong}
\end{equation}
where,  ${\cal C}_{\bf m}$ is the sky error covariance, given in general by Eq.~\ref{eq:destMapError}, and which for our purpose in this section will be approximated as either diagonal or block-diagonal for the unpolarized and polarized detectors respectively as given by the first term on the right-hand side ({\it rhs}) of Eq.~\ref{eq:destMapError}. 

The first term on the rhs of Eq.~\ref{eq:offCovLong} describes the offset variance derived assuming that the sky signal is known. The effect of the marginalization over the unknown sky performed in the presence of the crossings between the rings is then quantified by the second term. We emphasize that no correction is explicitly applied in the above equation to account for the singularity of the problem (see Sect.~\ref{sec:polkapixOff}). Though this is sufficient for the qualitative discussion presented here, it does have to be treated in the actual numerical work. In the following, we discuss separately two specific cases of interest: one assuming unpolarized total power detectors and the other polarization-sensitive ones. For simplicity, we assume the noise variance per sample, $\sigma_t^2$, to be time-independent and the same for all considered detectors.

\subsubsection{Total power detectors}
\label{sec:covmat_temp}

\begin{figure*}[h!]
	\center
	\includegraphics[width=0.32\textwidth]{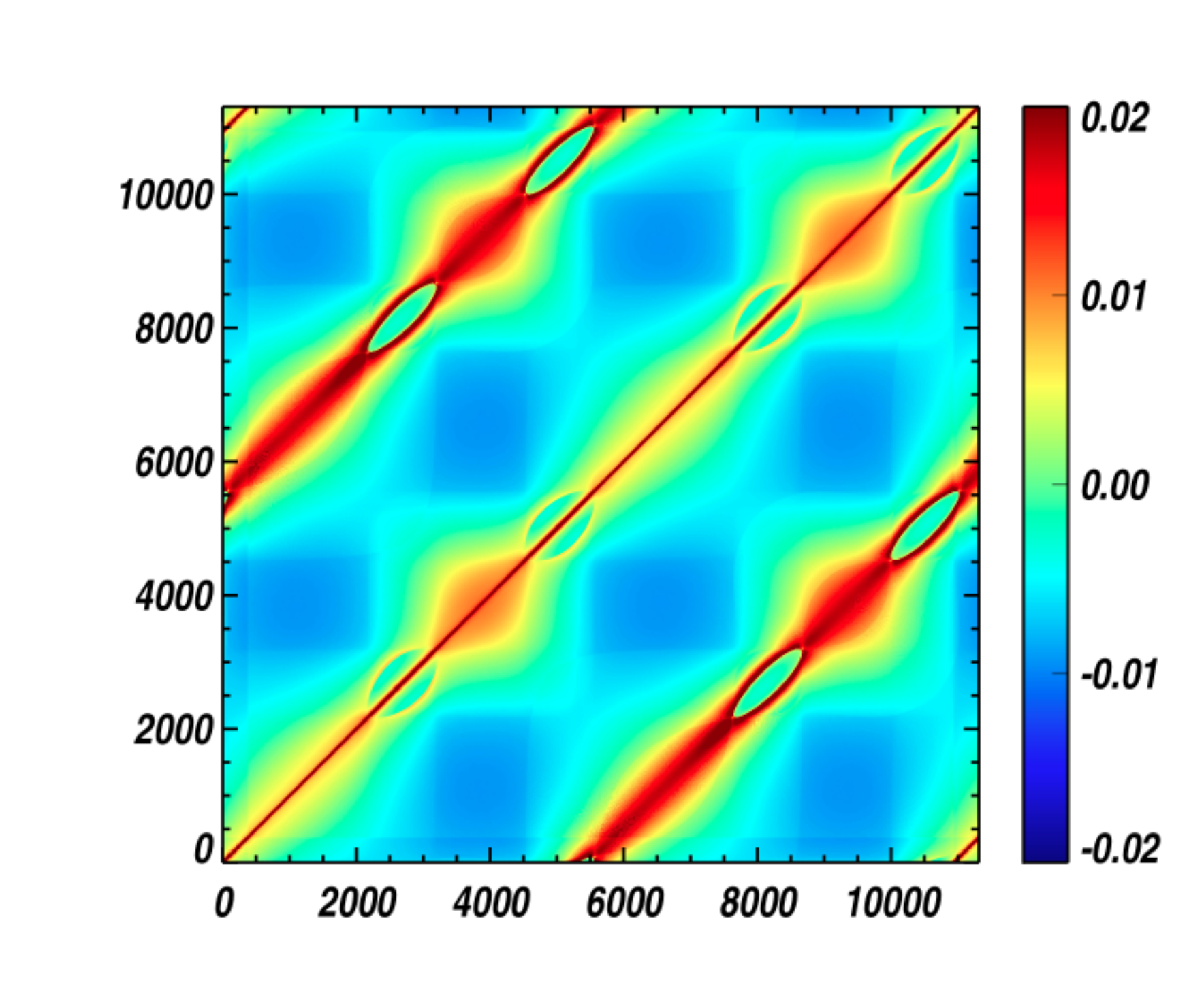}
	\includegraphics[width=0.32\textwidth]{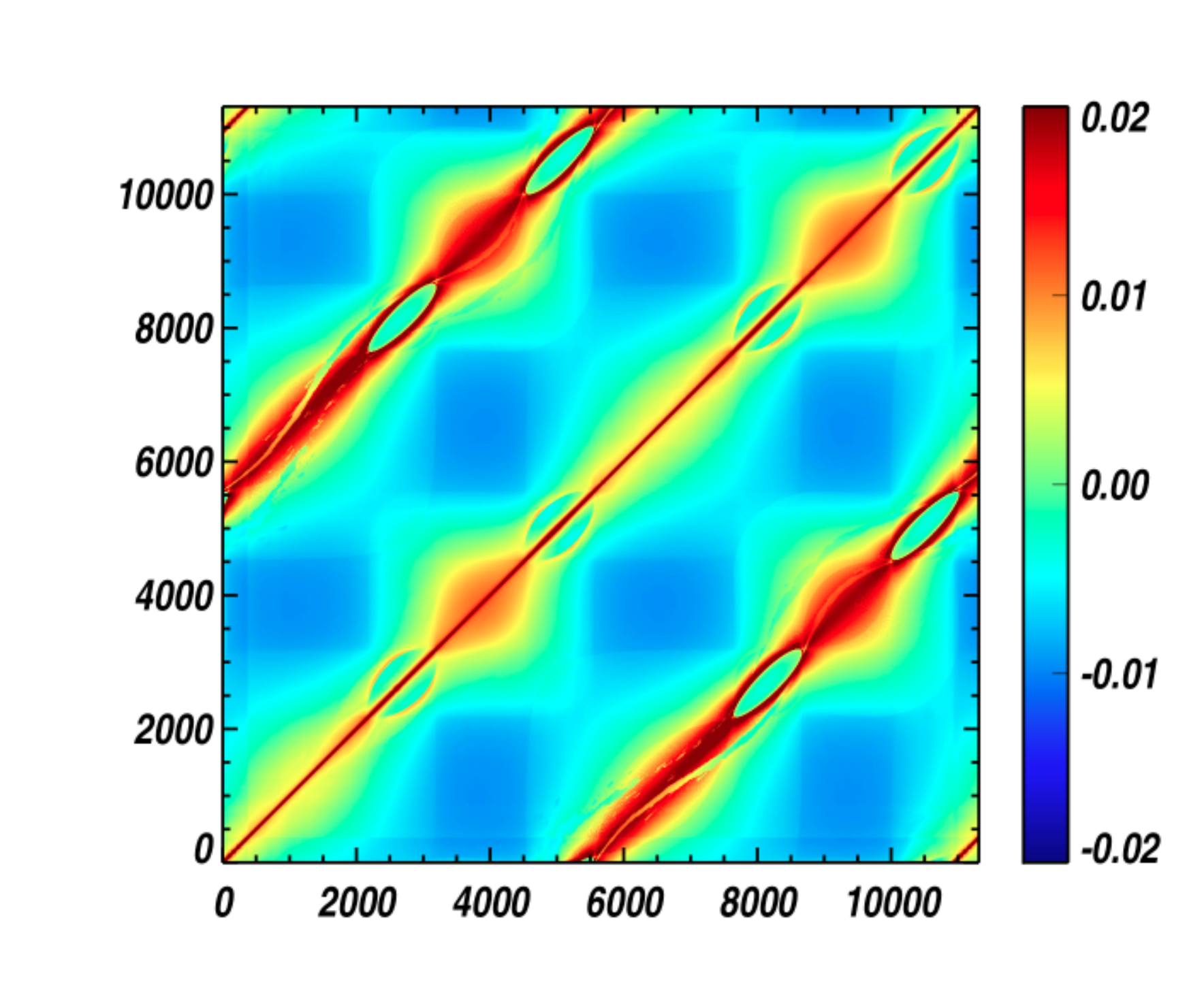}
	\includegraphics[width=0.32\textwidth]{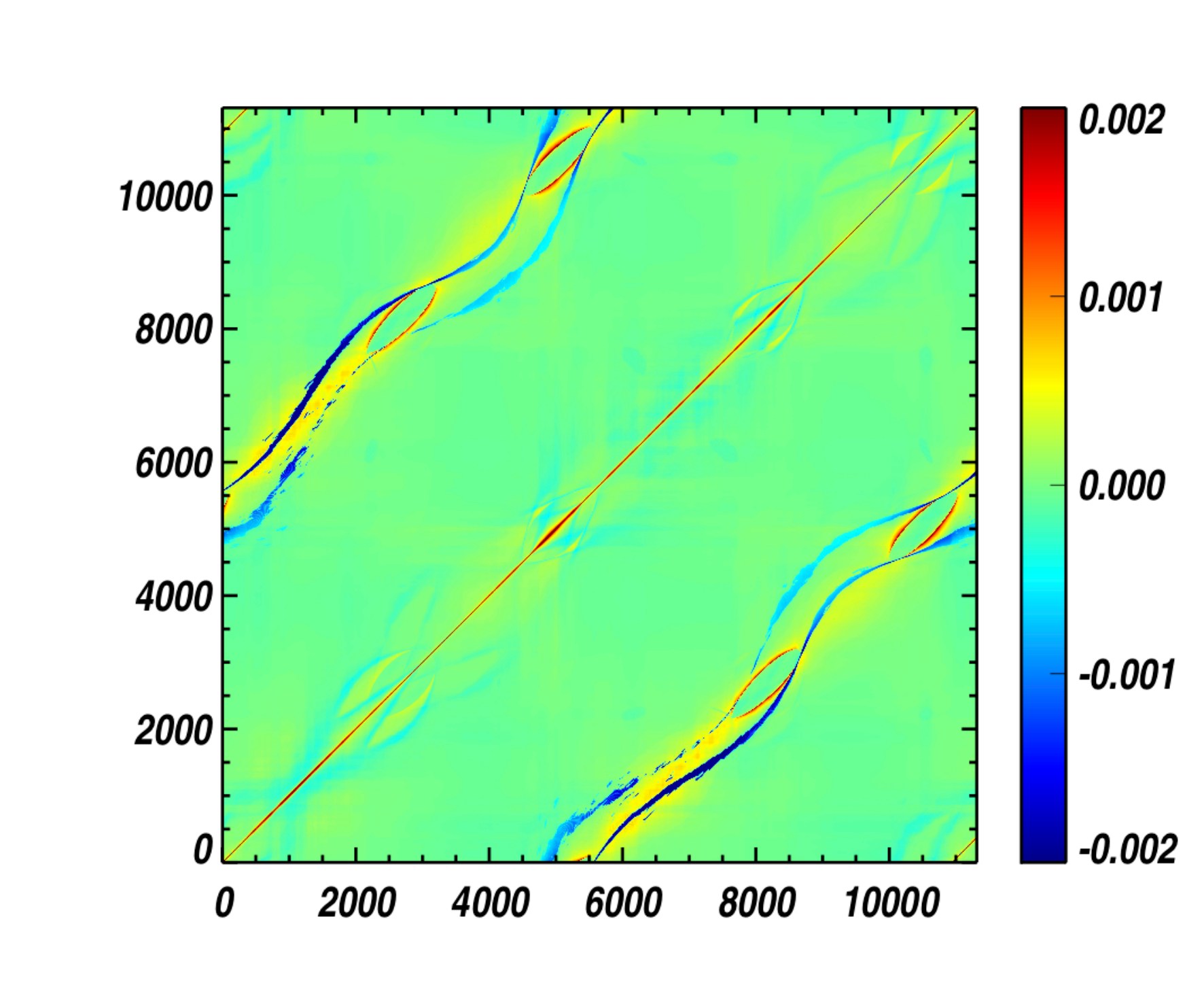}
	\caption{Offset covariance matrix without ({\it left}), with a Galactic mask ({\it middle}) and the difference ({\it right}).} 
\label{fig:covmat_mask}
\end{figure*}

\begin{figure*}[h!]
	\center
	\includegraphics[width=\textwidth]{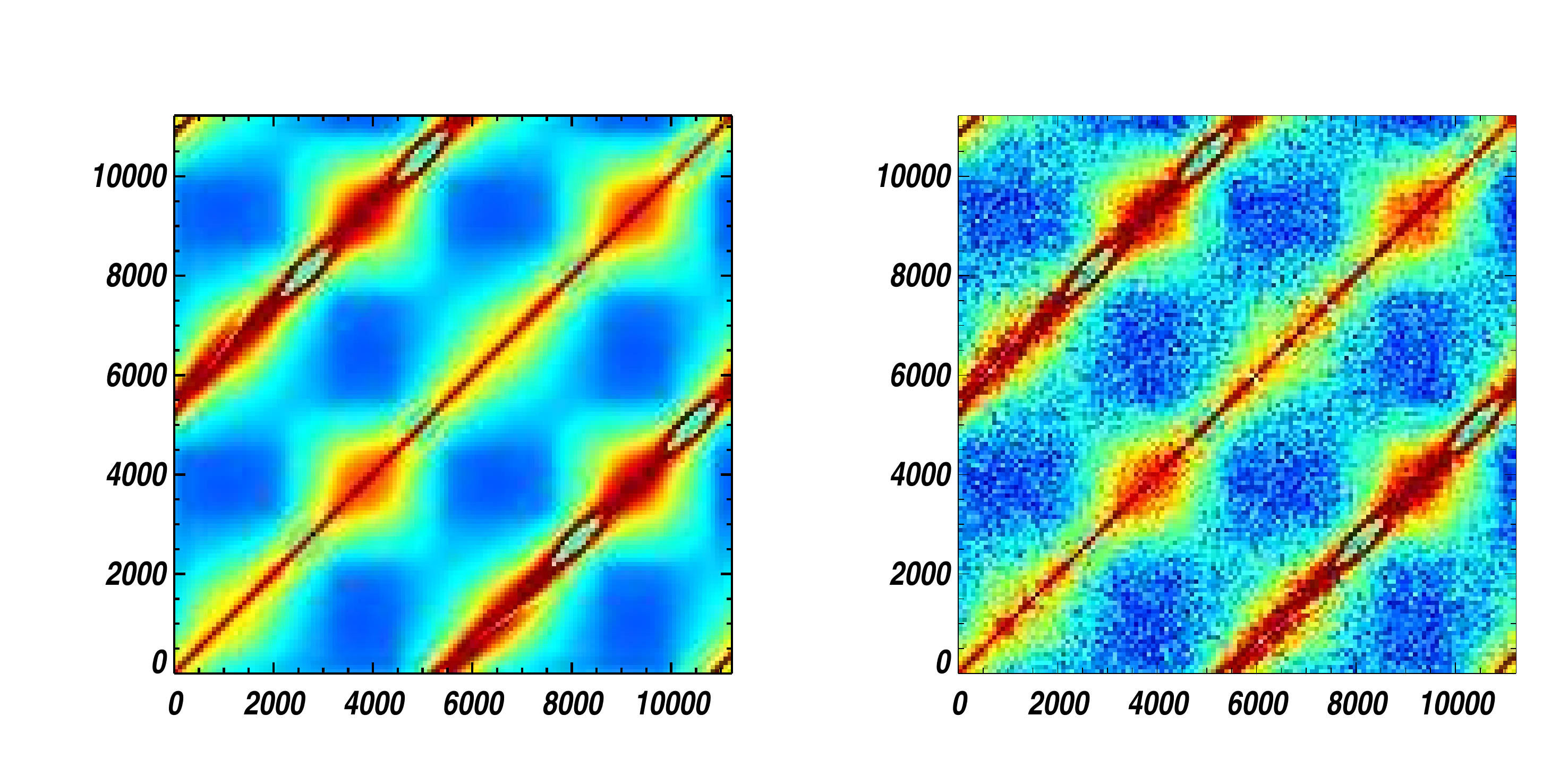}
	\caption{Binned offset covariance matrices. {\it left:} {\sc Polkapix} estimate, {\it right:} covariance for \nsimu~MC. The bin width is 100~rings.}
\label{fig:covmat}
\end{figure*}

For total power detectors, the entries of the sky pointing matrix, ${\mathbf A}$, are either $0$ or $1$, with the indices of the non-zero elements, $(t,\ p)$, defining which pixel $p$ has been observed at the time, $t$. The matrix, ${\mathbf C_m}$ is  approximately diagonal with the elements equal to the estimated noise level for each pixel. Elements of the matrix, ${\bf \Gamma}^t \, {\bf N}^{-1} \, {\bf A}$  are then given by a number of times each pixel was observed within each offset baseline, i.e., a ring, weighted by the sample noise.  In this special case, we can therefore rewrite Eq.~\ref{eq:offCovLong} as
\begin{equation}
	\left[{\cal C}_{\bf x}\right]_{(d,r) (d',r')} \simeq   \frac{\sigma_t^2}{n_{obs}(r,d)} \left( \delta^K_{r r'}\,\delta^K_{d d'} +  \frac{1}{n_{obs}(r',d')}\,\sum_{p\in r \cap r'}\,\frac{n_{obs}(p, r, d) n_{obs}(p,r', d')}{n_{obs}(p)}\right).
\label{eq:offCovTemp}
\end{equation}
Hereafter we use $n_t$, $n_r$ to denote the total number of samples and rings per detector; $n_p$ the total number of  pixels observed by all detectors; $n_{obs}(p,r,d)$ the number of observations of a pixel $p$ within a pointing period (ring), $r$, for a detector, $d$, where $n_{obs}(p)\equiv \sum_{r,d}\,n_{obs}(p, r, d)$ is the total number of its observations; and $n_{obs}(r, d)$ the number of samples in a pointing period $r$.
Moreover, $r$ and $r'$ denote two rings corresponding to the pointing periods of two detectors, $d$ and $d'$, respectively, and a symbol $p \in r\cap r'$ refers to the pixels observed during both pointing periods, $r$ and $r'$.  We note that the second term on the {\it rhs} is always non-negative and vanishes if $r\cap r' = \emptyset$. It therefore increases the offset determination uncertainty and leads to a correlation, rather than an anti-correlation, of the offsets. This reflects that the offset determination precision is always slightly lower whenever the sky signal has to be estimated simultaneously. This has an easy intuitive explanation: if one of two offsets corresponding to two rings, which cross on the sky, turns out to be overestimated, the sky signal in pixels common to both rings will be on average underestimated, and the offset of the second ring will tend to compensate for this and therefore be overestimated as well. Nevertheless, the anti-correlations are unavoidably present in the offset correlation matrices. This can be seen in Figs.~\ref{fig:covmat_mask} and~\ref{fig:covmat} and arises because of the condition we impose on the sum of all offsets in order to break the problem singularity. This effectively requires that the sum of all the elements in each matrix column or row vanishes, hence that at least some, or more typically many, elements of the correlation matrix are indeed negative.

We first consider the case with $r=r'$ and $d=d'$. Equation~\ref{eq:offCovTemp} then reads
\begin{equation}
	\left[{\cal C}_{\bf x}\right]_{(d,r) (d,r)} \simeq  \frac{\sigma_t^2}{n_{obs}( r, d)} \left( 1 +  \frac{1}{n_{obs}(r,d)}\,\sum_{p\in r }\,\frac{n^2_{obs}(p, r, d)}{n_{obs}(p)}\right).
\label{eq:offCovTemp1}
\end{equation}
Given that typically $n_{obs}(r,d)\sim n_t/n_r \gg 1$, the second term in the parenthesis is strongly suppressed. However, the magnitude of the sum over the pixels observed in the period, $r$, can be large enough to compensate for this, rendering this term important. To illustrate this, we first observe that
\begin{equation}
	\sum_{p\in r}\, n_{obs}(p, r,d) = n_{obs}(r,d) \sim  \frac{n_t}{n_r},
\end{equation}
and consequently that the sum in Eq.~\ref{eq:offCovTemp1} is larger, whenever $n_{obs}^2(p,r,d)$ varies more rapidly from a pixel to a pixel along the ring, or, in other words, whenever the distribution of the observation within the pointing period, $r$, is more inhomogeneous. In contrast, whenever both $n_{obs}(p, r, d)$ and $n_{obs}(p)$ are homogeneous, thus pixel-independent, the second term is given as
\begin{equation}
	\frac{1}{n_{obs}(r,d)}\,\sum_{p\in r }\,\frac{n^2_{obs}(p, r, d)}{n_{obs}(p)} \sim \frac{n_p}{n_r n^2_p(r,d)} \simlt \frac{1}{n_p( r, d)} \ll 1,
\label{eq:caseHom}
\end{equation}
where $n_p(r,d)$ is the number of pixels observed in the period, $r$. The offset covariance is then close to its minimum, where it is dominated by the first term, with the second term contribution nearly negligible thanks to an implicitly implied perfectly linked network of rings on the sky.

\begin{figure*}[h!]
	\center
	\includegraphics[width=0.49\textwidth]{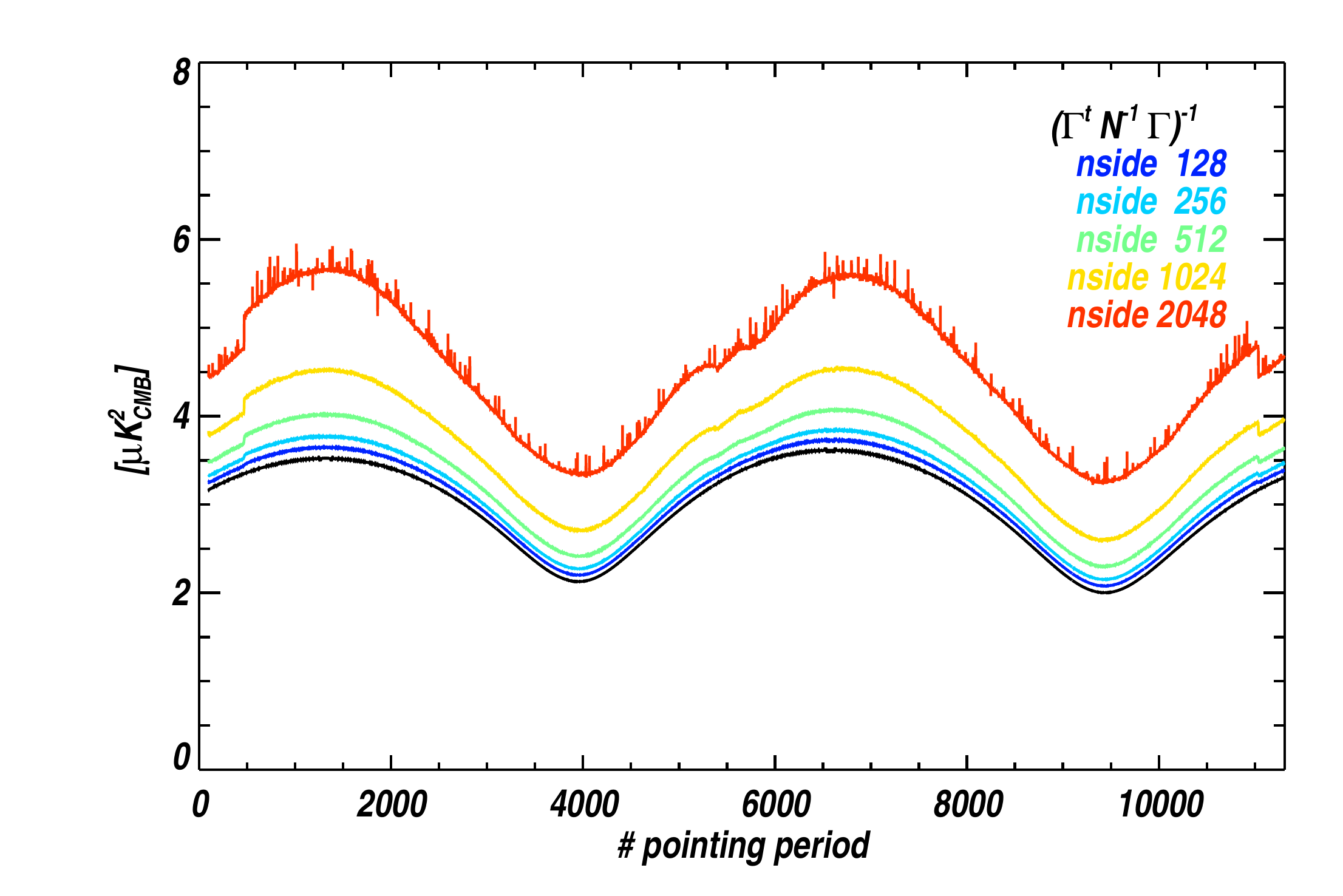}
	\includegraphics[width=0.49\textwidth]{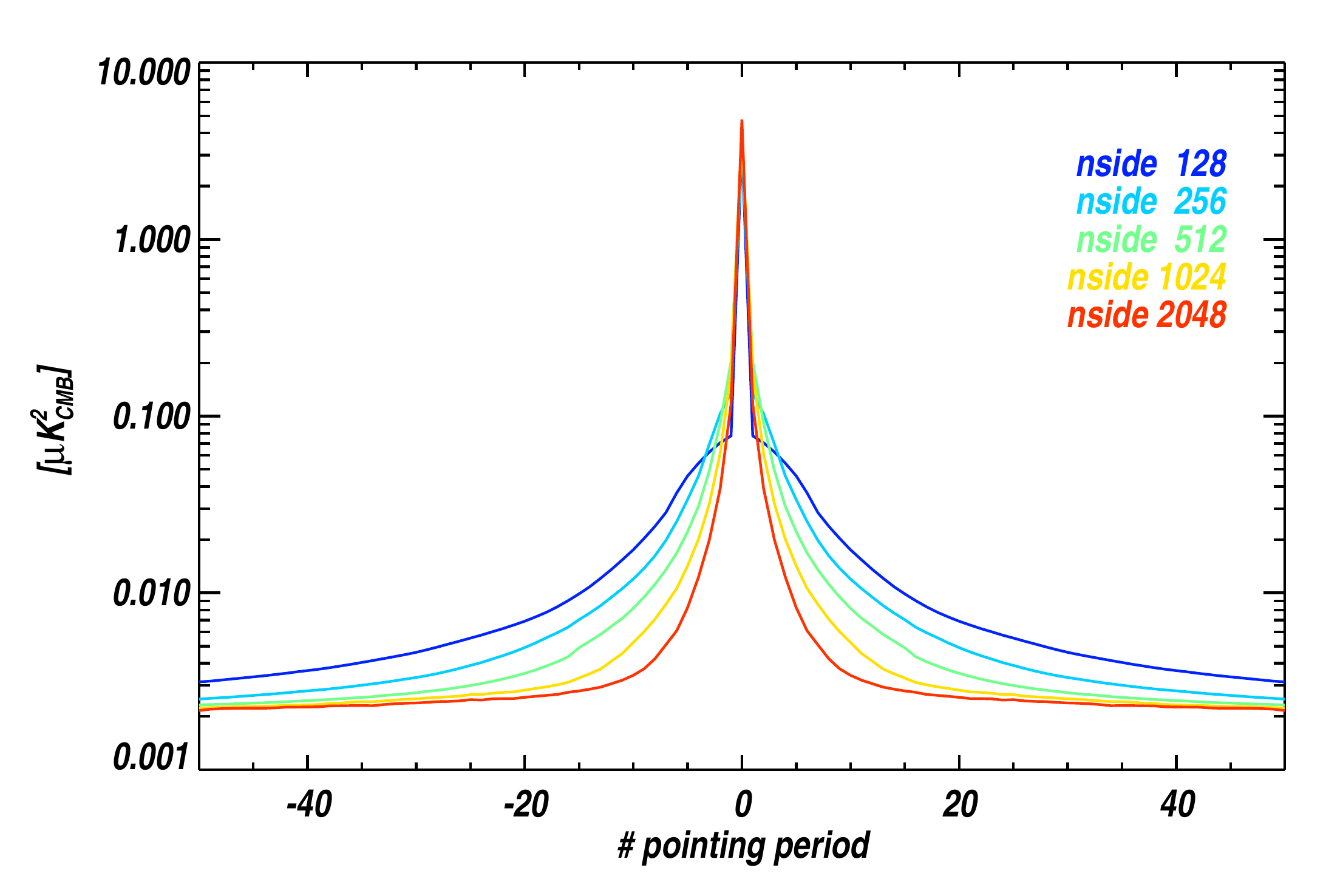}
	\caption{{\it Left:} Diagonal of the offset covariance matrix for various internal sky resolutions used for  the offset recovery. Black line shows the first term of Eq.~\ref{eq:offCovTemp}. {\it Right:} Offset covariance matrix profiles centered on the diagonal showing the ring-to-ring covariance for various internal sky resolution.}
\label{fig:covmat_nside}
\end{figure*}

At the other extreme, all samples of the ring $r$ are located in a single pixel, $p$. The second term is then roughly given by
\begin{equation}
	\frac{1}{n_{obs}(r,d)}\,\sum_{p\in r }\,\frac{n^2_{obs}(p, r, d)}{n_{obs}(p)} \sim \frac{n_t}{n_r n_{obs}(p)} \simlt 1,
\end{equation}
as $n_{obs}(p)  \ge n_{obs}(p,r,d) = n_t/n_r$ and the equality  is realized whenever $p$ is  observed only within the considered pointing period. We thus conclude that the second term, though typically smaller, may be comparable in magnitude to the first one. The important consequence of these considerations is that whenever the inhomogeneity of pixel observations for some pointing period increases the diagonal elements of the offset-offset covariance matrix tend to increase. This explains the results displayed in Fig.~\ref{fig:covmat_nside}, which display a noticeable increase in the offset variances as a result of a decrease in pixel size, which in turn enhances the observation inhomogeneity. This has important practical implications when selecting an appropriate pixel size during the offset determination step (see Sect.~\ref{sec:destriping_syste}). We note that for large pixels the second term is unimportant to the observation distribution among the pixels reaching the limit as discussed earlier in the context of Eq.~\ref{eq:caseHom}.

For two different rings, i.e., $r \ne r'$ or $d \ne d'$,  the first term in the expression for the covariance is absent and therefore the latter is determined by the second term. Its magnitude is now driven by two effects, the first one -- as in the previous case -- related to the observations inhomogeneity, which in general tends to increase the strength of the correlations, and the second one related to the number of pixels in common between the two rings. The latter effect is however the one that drives the large-scale appearance of the offset correlation matrix, Fig.~\ref{fig:covmat_mask}. Consequently, the matrix is diagonal dominated with the correlations decaying away from the diagonal as the number of crossing pixels between the two rings decreases. The decay rate as well as the auto-correlation change with the ring number as does the Planck scan pattern, during which the satellite spin axis follows a cycloid in the Ecliptic coordinates. Consequently, the width of the diagonal band is found to vary along the diagonal. As Planck observes a full sky in six months of observations, with the end of the six month corresponding roughly to ring $5000$, the overlaps between the rings separated by $\sim 5000$ are again enhanced  leading to a significant level of the corresponding correlations. This again happens for the ring separation of on the order of $10000$, when the third full-sky survey starts. We note that the pattern of sky scanning changes between the first and second survey, owing to the different satellite tilt at which it is preformed, and consequently the offset correlations at this lag are somewhat more smeared out than those around the main diagonal. As the third scan follows the first nearly exactly, the correlations at the lag of $\sim 10000$ are expected to resemble those at the diagonal very closely as indeed already seen to some extent in the figure.

Owing to the two competing effects mentioned above, the impact of the pixel size on the off-diagonal correlations is more complex than on the offset auto-correlations as they tend to change the correlation amplitude in an opposite way. Hence, though the correlation length will decrease when any pixel is shared on average by fewer rings, the changes in the correlation amplitude will depend on the fine details of the scanning strategy and in general will be neither simple nor easy to predict. This is indeed confirmed by the tests presented in the right panel of Fig.~\ref{fig:covmat_nside}.

The contribution of the second term in Eq.~\ref{eq:offCovTemp} is sensitive to the pixels that are retained for the offset determination. This is true in particular when only a few pixels are shared between two rings, or whenever well-observed pixels are removed. In this context, the pixel selection is vitally important to practical applications as discussed in more detail in Sect.~\ref{sec:destriping_syste}. Here we only illustrate it in Fig.~\ref{fig:covmat_mask} by contrasting two covariance matrices, obtained with all sky pixels included (left panel) or with those corresponding to the discarded Galactic plane (middle).

For two rings that are different on the sky, whether they correspond to pointing periods of a single or two different detectors is completely irrelevant. This is because of our assumptions about lack of time-domain noise correlations and the same value of the noise variance. This, together with all the detectors being assumed to scan the sky in the same way, explains the periodic structure of the four detector offset covariance shown in Fig.~\ref{fig:covmat_multibolo}. We indeed see that all the off-diagonal blocks are essentially identical, while the only difference between any of the diagonal and any of the off-diagonal blocks is caused by their diagonal values and is a result of the non-vanishing contribution to the first term in the case of the diagonal blocks. Our assumptions about the scanning imply that $n_{obs}(p,r,d)$ is independent of the detector number, rendering the off-diagonal blocks symmetric with respect to their diagonal.

We also checked our calculation of the offset covariance matrix from {\sc Polkapix} using \nsimu~Monte-Carlo simulations of pure white noise. We simulated a simplified sky with CMB and Galactic emissions for a single {\it Planck}-HFI detector at 217~GHz. We then generated \nsimu~white noise realizations. For each simulation, we estimated the offsets using a 5\% Galactic mask. We found a very good agreement between the MC standard deviation and the covariance matrix as shown in Fig.~\ref{fig:covmat}.

\subsubsection{Polarization-sensitive detectors}

\begin{figure}[h!]
	\center
	\includegraphics[width=240pt,height=230pt]{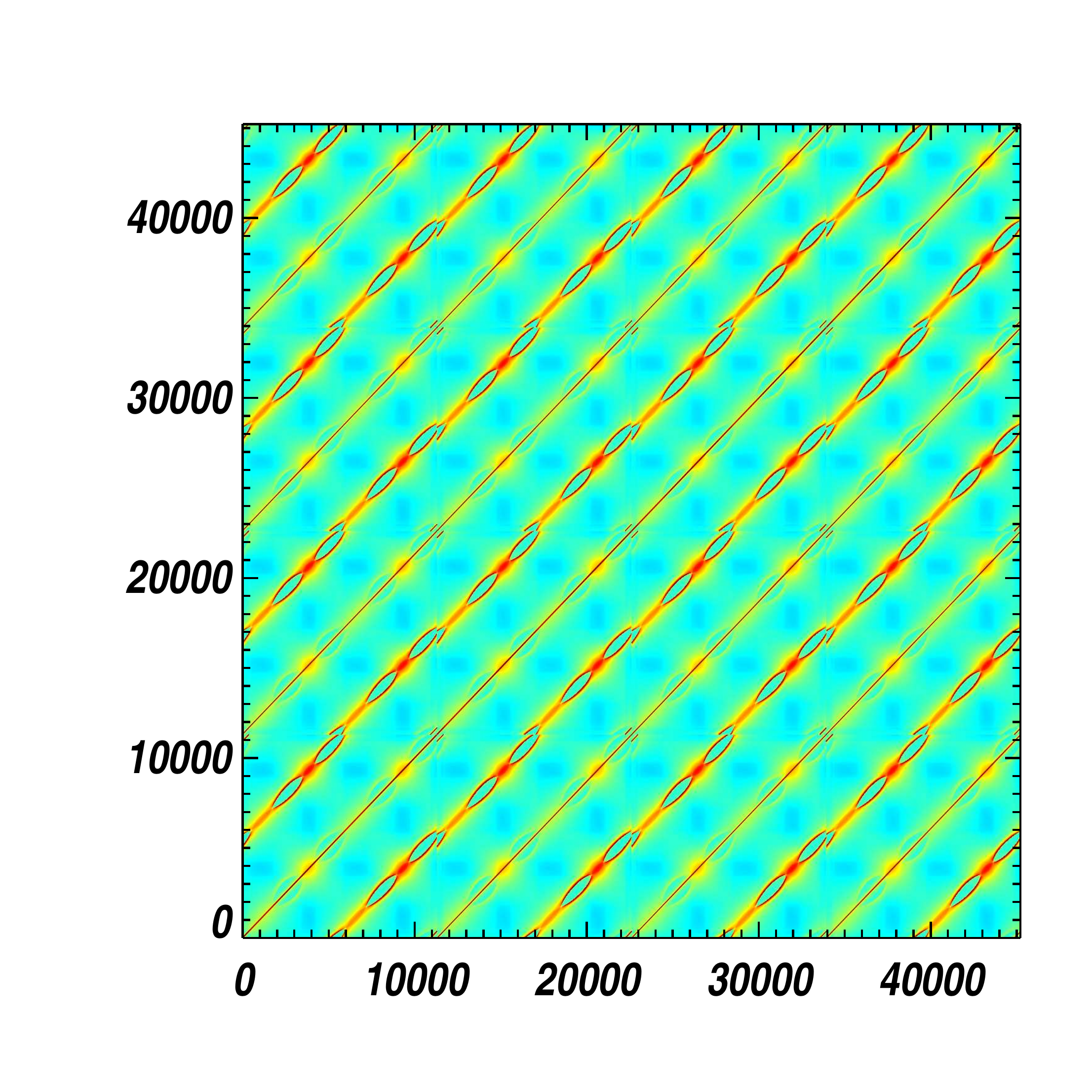}
	\includegraphics[width=270pt,height=230pt]{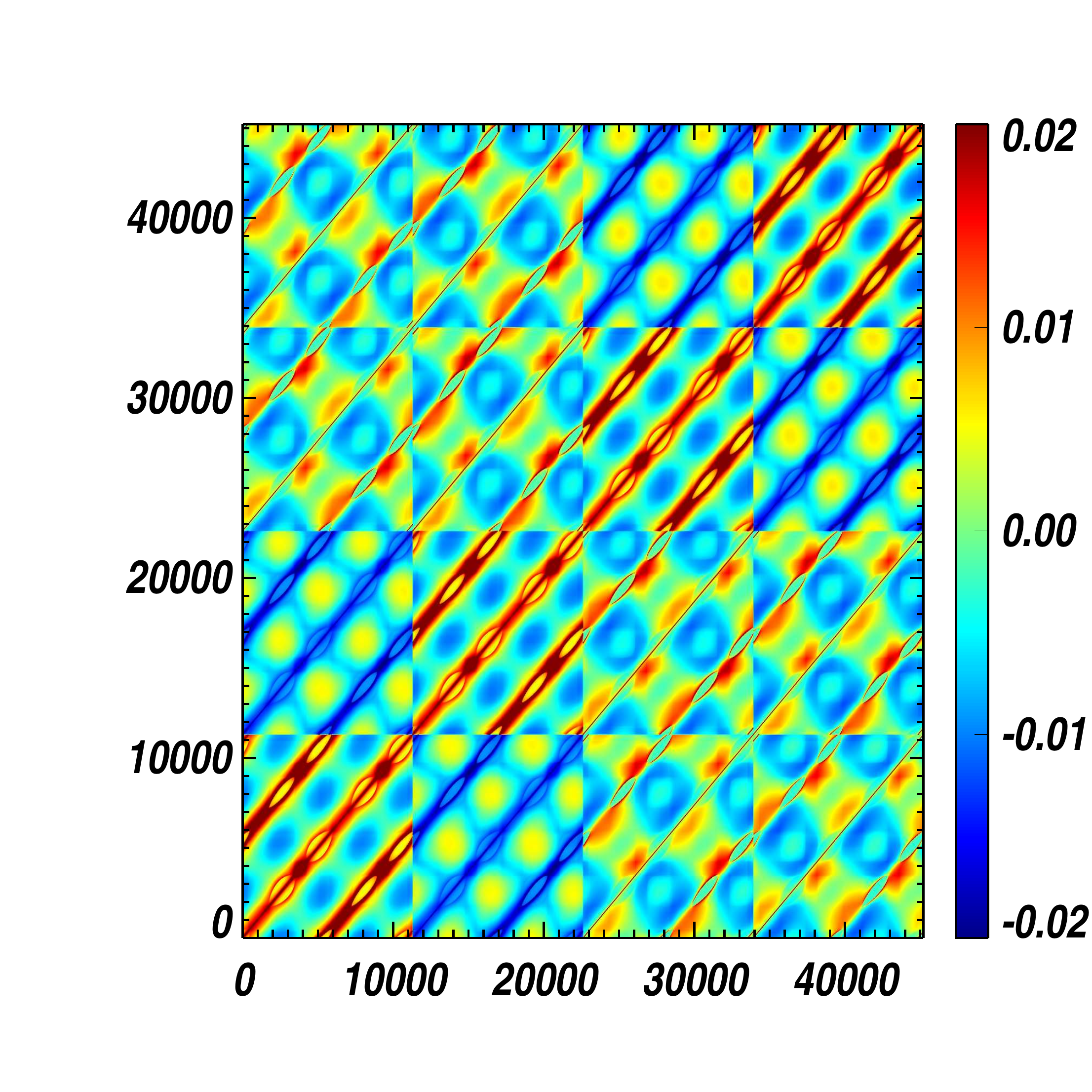}
	\caption{Offset correlation matrix for four polarization-sensitive bolometers at 143GHz without ({\it left}) and with ({\it right}) polarization reconstruction in the destriping.} 
\label{fig:covmat_multibolo}
\end{figure}

For polarization-sensitive detectors, the pointing matrix, $\mathcal{A}$, is more complex, Eq.~\ref{eq:polDataModel}, with elements, that can be both positive and negative. The sky covariance can be approximated, as before, by the first term of Eq.~\ref{eq:destMapError}. This would in general render a block-diagonal structure with each block of a size $3\times3$ characterizing single-pixel correlations of three Stokes parameters. However, in our case (see Sect.~\ref{sec:simulations} and in particular Eq.~\ref{eq:pixNoise}) where the data of four detectors are analyzed simultaneously, it can again be treated as diagonal with the diagonal elements corresponding to $Q$ and $U$ parameters always being a factor $2/\varepsilon^2$ different than those for the total intensity, $I$. Hereafter, for brevity we omit the polarization efficiency $\varepsilon$. Assuming as before that the white noise is stationary over the course of the entire set of observations for which an rms value, $\sigma^2_t$,  is identical for all detectors, we can now specialize Eq.~\ref{eq:offCovLong} to the case at hand obtaining
\begin{eqnarray}
	\left[{\cal C}_{\bf x}\right]_{(d,r) (d',r')}  
	\simeq
	\frac{\sigma_t^2}{n_{obs}(r,d)} \Biggl\{ \delta^K_{r r'}\,\delta^K_{d d'} +
	&& \frac{1}{n_{obs}(r',d')}\,\sum_{p\in r \cap r'}\, \frac{n_{obs}(p, r, d) n_{obs}(p,r', d')}{n_{obs}(p)} \; \Biggl.\Bigl. \nonumber\\
	&& \Biggl.\Bigl.  \Bigl[ 1+ 2\,\left(\left\langle\cos 2\psi\right\rangle_{(p, r, d)} \left\langle\cos 2\psi\right\rangle_{(p, r', d')}
	 +  \,\left\langle\sin 2\psi\right\rangle_{(p,r,d)} \left\langle\sin 2\psi\right\rangle_{(p,r',d')}\right)\Bigr]
	\Biggr\}.	
\label{eq:offCovPol}
\end{eqnarray}
Here,
\begin{equation}
\left\langle\cos 2\psi\right\rangle_{(p,r,d)}  \equiv  \frac{1}{n_{obs}(p,r,d)}\,\sum_{t\in p \cap r}\,\cos 2\psi_t^d.
\end{equation}
We note that for each ring, $r$, and each detector, $d$, the averages above are done  separately and therefore the arguments used in simplifying Eq.~\ref{eq:pixNoise} do not apply here and the terms with cosines and sines do not vanish trivially in the Planck case. In contrast, they have an important and easily discernibly impact on the offset covariance matrix as shown in Fig.~\ref{fig:covmat_multibolo}.

For Planck scans, the angle between the polarizer orientation and the celestial coordinate system for a given pixel $p$ and during a ring $r$  does not change significantly from one to another pixel crossing as they are fixed with respect to the scanning direction. The averages of the trigonometric functions are therefore well approximated by the cosines and sines of this specific angle. Equation~\ref{eq:offCovPol} can then be simplified yielding
\begin{eqnarray}
	\left[{\cal C}_{\bf x}\right]_{(d,r) (d',r')}
	\simeq 
	\frac{\sigma_t^2}{n_{obs}(r,d)} \Biggl\{ \delta^K_{r r'}\,\delta^K_{d d'} +  \frac{1}{n_{obs}(r',d')} 
	&& \sum_{p\in r \cap r'}\, \frac{n_{obs}(p, r, d) n_{obs}(p,r', d')}{n_{obs}(p)} \Biggr.  \nonumber\\
	&& \Bigl[ 1+  2\,\cos \Bigl( 2(\psi(p, r, d)-\psi(p, r', d'))\Bigr)\Bigr]
	\Biggr\}.
\label{eq:offCovPol1}
\end{eqnarray}

Consequently, for any single ring, i.e., $r=r'$ and $d=d'$, the extra terms due to polarization are always positive, and, twice as large as the third, angle-independent term on the rhs. For the two corresponding rings belonging to two detectors in the same horn, e..g, with the relative polarizer orientation at $90$ degrees, the third term is equal to $-2$ forcing the corresponding element of the offset covariance to be negative. If the two detectors above are chosen from two different horns, i.e., their polarizers are either at $45$ or $135$ degrees with respect to each other, the corresponding rings of the detectors are again positively correlated. Moreover, in the latter case and for two different rings, $r$ and $r'$ the cosine function in Eq.~\ref{eq:offCovPol1} turns into a sine and the contribution is asymmetric with respect to an exchange of $r\leftrightarrow r'$ giving rise to an asymmetry of the corresponding off-diagonal blocks. This is not so for the same horn detector blocks as the correlations are given by a cosine function, symmetric with respect to ring exchanges. All these observations and in particular a dichotomy between identical and different horn detectors are indeed confirmed by our numerical calculations, Fig.~\ref{fig:covmat_multibolo}.
 
As in the unpolarized case discussed earlier, we can track the pattern of each block to the specific, single-detector scan features, such as the beginning of the second ($r\sim 5,000$) and the third ($r\sim 10,000$) full sky survey, leading to an overlap between their respective rings with those of the previous surveys and thus to an enhancement in the strength of the correlations.
 
\subsubsection{Offset covariance -- a recap}
We briefly summarize all salient points of the discussion presented above. We have found that in the Planck case the offset covariance matrix is in general non-diagonal and has a rather complex structure, major features of which are due to the pattern of the Planck scan. The computation of this matrix requires rather lengthy and involved numerical calculations, although quick intuitive insights can be obtained in a semi-analytic way \citep[see also, e.g.,][]{stompor04,efstathiou07}. Those can be not only useful in devising new, efficient scanning strategies but also provide a useful cross-check of the numerical results. We have also tested the latter by means of extensive Monte Carlo simulations and found good agreement.

We have demonstrated that the offset variance and their correlations depend on the observation (in)homogeneity and thus on the assumed pixel size and the number of crossing points between the rings, i.e., sky areas covered by the experiment within a single pointing period. We have also found that the overall structure of the covariance depends on whether the detectors are polarization-sensitive, an effect that, for a Planck-like scanning, has to be accounted for if a high-precision offset determination is desired. This last conclusion seconds one of the conclusions obtained in Sect.~\ref{sec:destriping_syste}.

\subsection{Residuals from low frequency noise\label{sec:destriping_fknee}}

In the previous sections, we have described the systematic effects related to the destriping algorithm presented in Sect.~\ref{sec:polkapix}. In particular, we have shown that the pixelization effects linked to the strong variation in the sky signal within a pixel are negligible for the offset estimation. However, we have found that a significant level of residuals arise whenever the polarization signal is neglected or the resolution adopted for the estimation of the underlying sky signal is reduced.
In this section, we study the residuals for various levels of correlated noise. For each bolometer, we add to the simulated signal (containing CMB and Galactic emissions) $100$ noise realizations for three cases: pure white noise, correlated noise with $\alpha=2$ for two knee frequencies ($f_{knee} = 0.1$ and $0.01$~Hz), and $\alpha=1$ for a knee frequency of $0.1$~Hz. We reduce the systematic bias using a $5$\% internal Galactic mask and an internal sky resolution of $6.9$~arcmin (HEALPix nside = $512$), as shown in Sect.~\ref{sec:destriping_syste}.

\begin{figure*}[h!]
	\center
	\includegraphics[width=\textwidth]{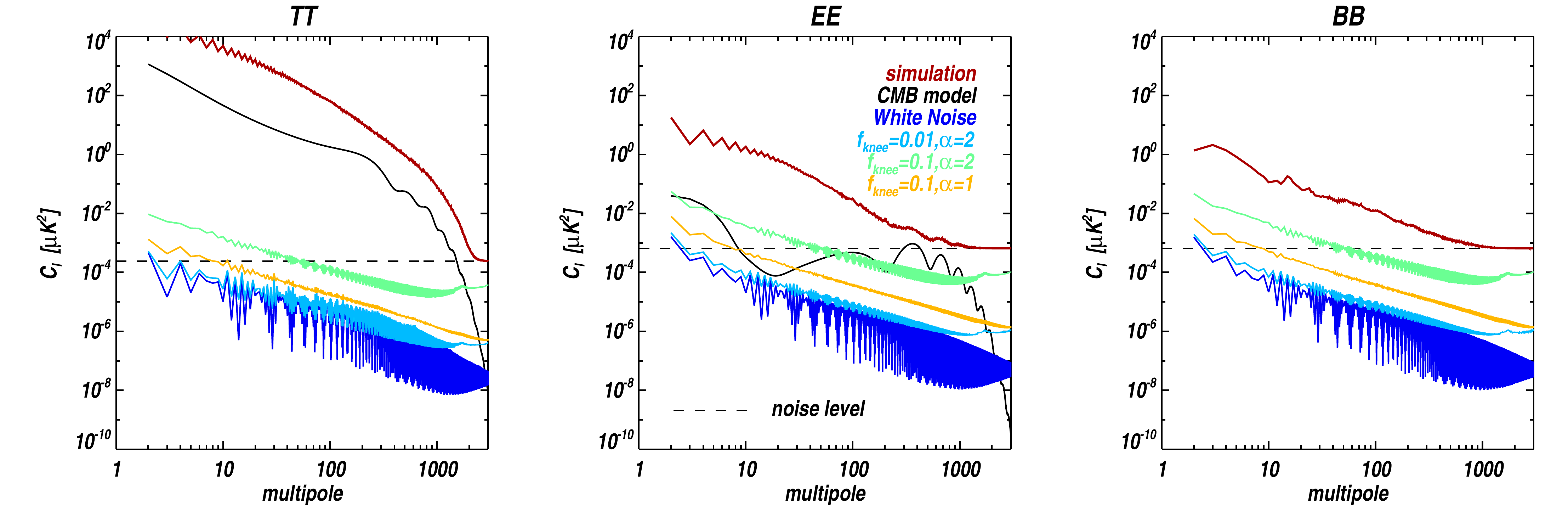}
	\includegraphics[width=\textwidth]{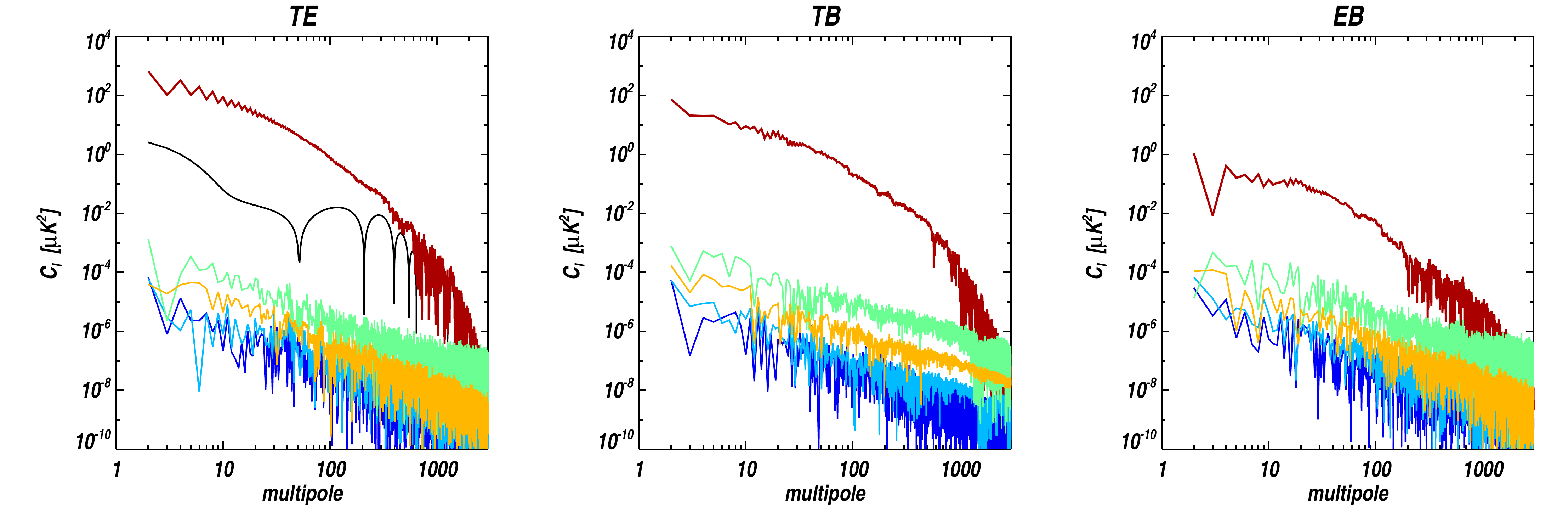}
	\caption{Residuals spectra after destriping the signal plus noise in the three cases of (1) pure white noise ({\it blue}), low frequency noise plus white noise with ($\alpha$=2,$f_{knee}$=0.1~Hz), (2) ($\alpha$=2,$f_{knee}$=0.01~Hz) and ($\alpha$=1,$f_{knee}$=0.1~Hz) compared to the simulation power spectra ({\it red}), and (3) CMB WMAP five-year fiducial model ({\it black}) on 100 simulations. {\it Upper line, from left to right:} temperature, E-mode, and B-mode. {\it Bottom line, from left to right:} cross-correlation TE, TB, and EB.}
\label{fig:fknee_pol}
\end{figure*}

Figure~\ref{fig:fknee_pol} shows the power spectra of the residuals computed for the three selected cases in both temperature and polarization. The level of the residuals is determined by the level of low-frequency noise introduced into the TOD. Thus, residuals increase with both the knee frequency $f_{knee}$ and the slope $\alpha$. We show that low-frequency noise residuals are negligible for $f_{knee} \le 0.01$~Hz when $\alpha=2$. For a higher $f_{knee}$, a residual at low frequency biased the spectrum from a pure white-noise power spectrum at low multipoles ($\ell < 50$). In the case of $\alpha=1$, the residuals are lower by one order of magnitude.

Apart from the small deviations coming from the realistic scanning strategy used in this study, we found results that are fully compatible with the analytical predictions of \citet{efstathiou07}. In particular, we found that the power spectra of the residuals scale like $\sigma^2$ (where $\sigma$ is the white noise level in map domain) in a similar way to the noise power spectra itself. This implies that the results discussed in this section do not depend on the value of the white noise level but only on the characteristics of the correlated noise (through the $f_{knee}$ and $\alpha$ parameters of the noise Fourier spectrum).

\section{Calibration results}
\label{sec:calibration_results}

\subsection{Systematics}
\label{sec:calib_syste}

\begin{figure}[!h]
	\center
	\includegraphics[width=0.45\textwidth]{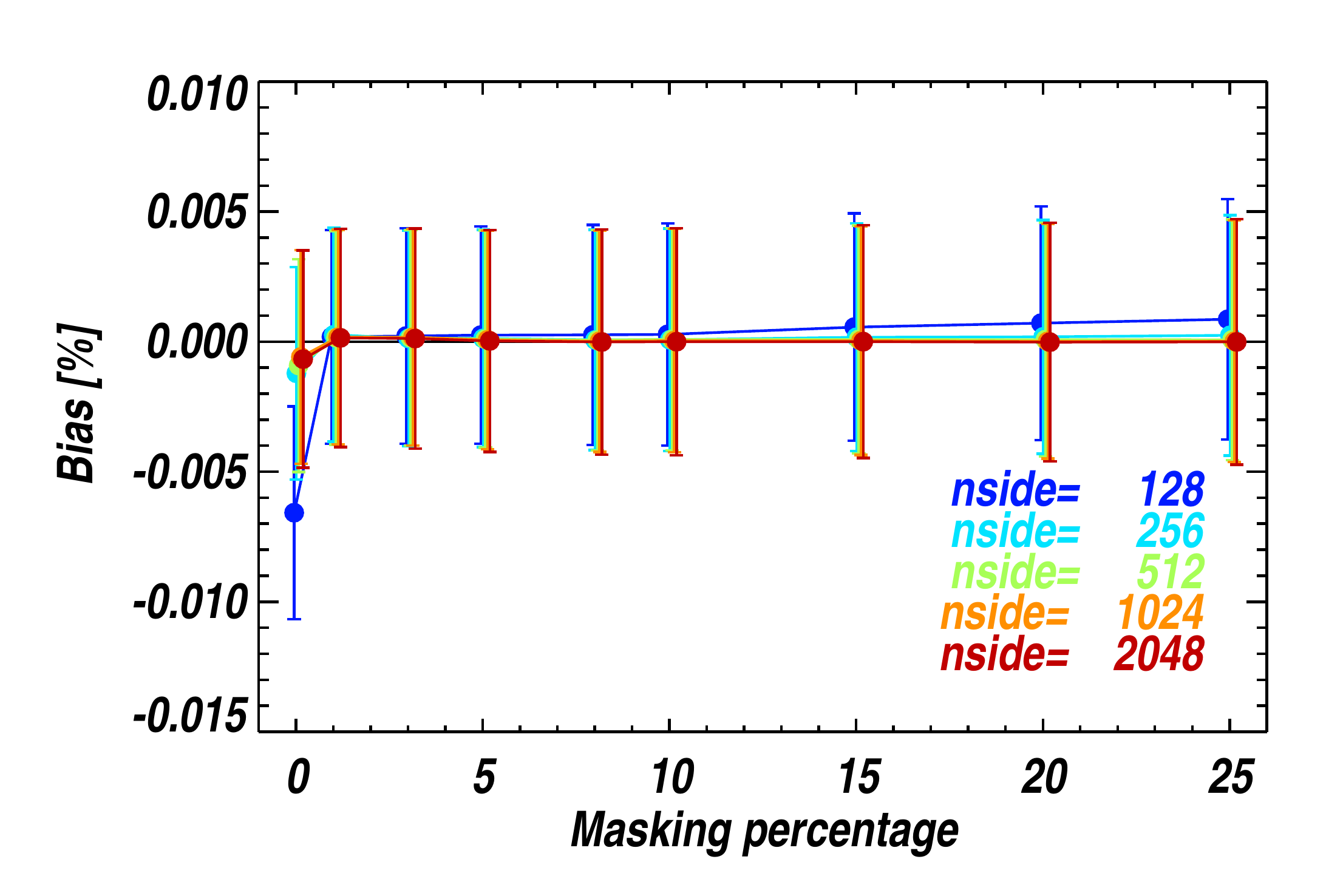}
	\caption{Evolution of the calibration bias due to the sky resolution with the masked fraction. Error bars are statistical estimates from {\sc Polkapix}.}
\label{fig:calib_bias_pixels}
\end{figure}

The calibration method is based on the assumption that when observing the same pixel of the sky at different times, the only variation in signal is due to the orbital dipole (assuming that the low-frequency noise is perfectly handled previously by the destriping). Any other effect leading to a difference between two different measurements of the sky power in one pixel, e.g. due to polarization because of a change in the bolometer orientation or a large signal variation inside the pixel) will result in a bias in the gain reconstruction.
We performed a simulation including only CMB and Galactic emission intensities and estimate the gain for several internal resolutions and various internal mask sizes. The masks are based on the gradient of the Galactic emission, which is the component inducing the strongest intra-pixel variations. We have found a bias smaller than $10^{-5}$ for all considered internal resolutions as soon as the 1\% most variable Galactic plane areas were masked. As shown in Fig.~\ref{fig:calib_bias_pixels}, the gain is biased only for the lowest resolution when the Galaxy is not masked at all ($g \leq 6\cdot10^{-5}$). We conclude that in the calibration case, intra-pixel variations induce very small systematics. To match the destriping settings, we fix hereafter the internal resolution to nside = 512. 

\begin{figure}[h!]
	\center
	\includegraphics[width=0.45\textwidth]{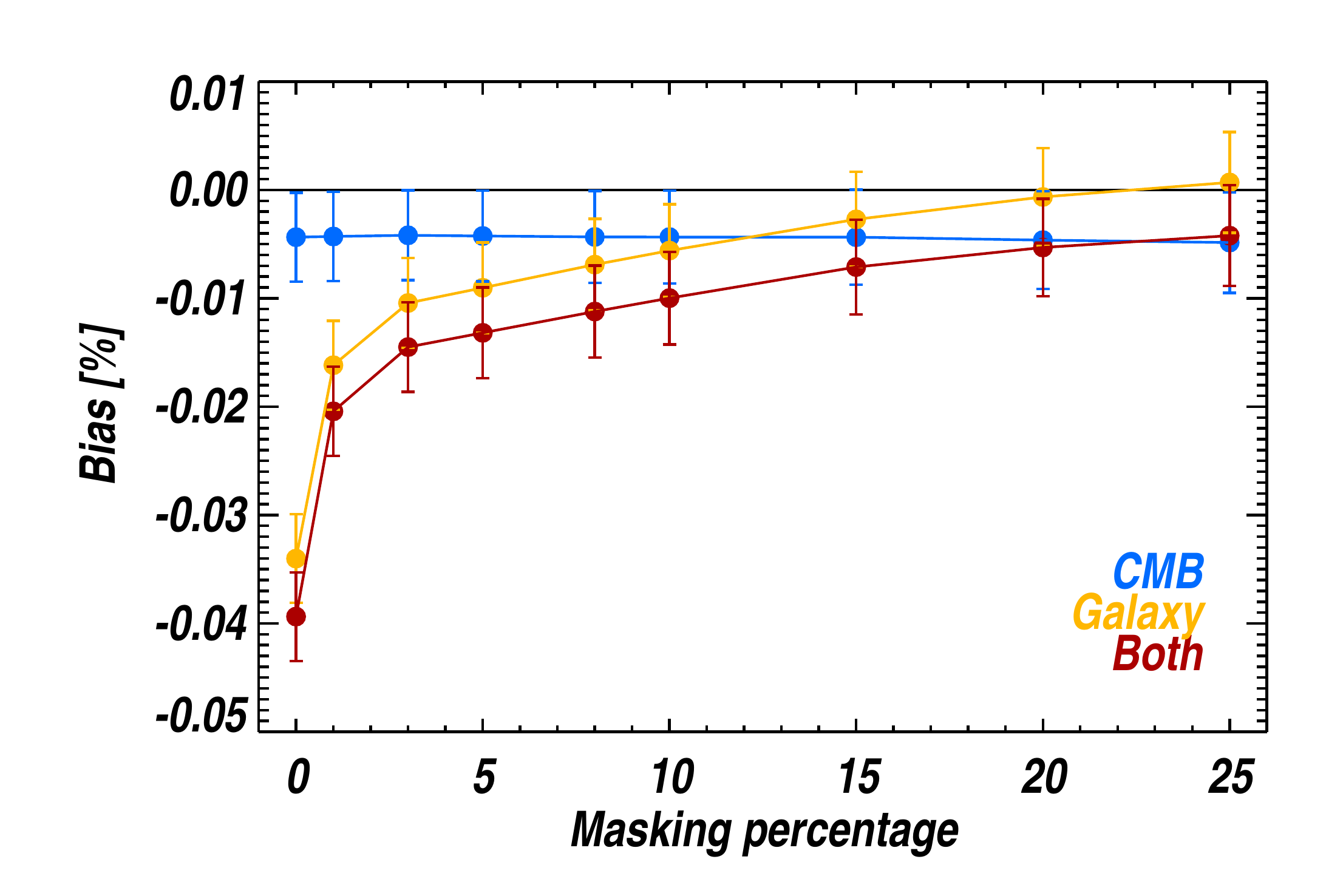}
	\caption{Evolution of the calibration bias due to the polarized signal and the masked fraction ({\it blue:} simulation of CMB; {\it orange:} simulation of Galactic emissions; {\it red:} simulation of CMB + Galaxy). Error bars are statistical estimates from {\sc Polkapix}.}
\label{fig:calib_bias_Polar}
\end{figure}

We also evaluate the bias in the gain reconstruction caused by the polarized sky signals from the Galaxy and CMB. We study three cases: CMB including intensity and polarization, Galactic emission with only polarized signal, and CMB + Galaxy (intensity and polarization). We reconstruct the gain applying several internal masks based on the intensity of polarization of the Galactic component.
Figure~\ref{fig:calib_bias_Polar} shows the bias of the reconstructed gain for the three simulations using each mask. We first note that the CMB polarization induces a bias (typically $\sim 5\cdot10^{-5}$) that is constant with respect to the percentage of sky masked but depends on the CMB realization and the bolometer characteristics. For a polarized Galactic signal, we find a stronger bias of up to $4\cdot10^{-4}$, which decreases with the mask size, reaching zero for a 20\% masked fraction. The statistical error bar increases with the size of the mask because of the smaller sky coverage.

\begin{figure*}[h!]
	\center
	\includegraphics[width=0.49\textwidth]{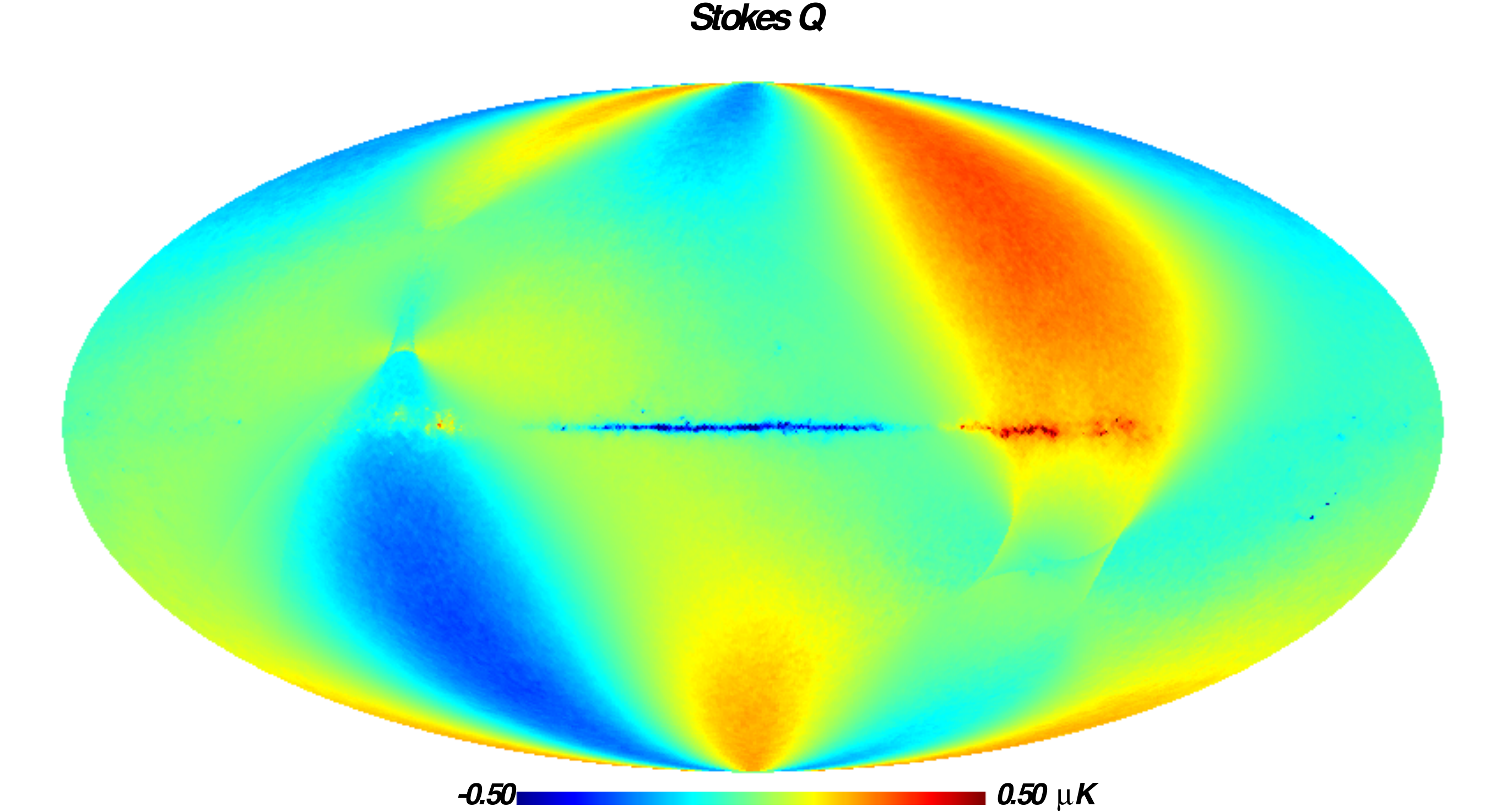}
	\includegraphics[width=0.49\textwidth]{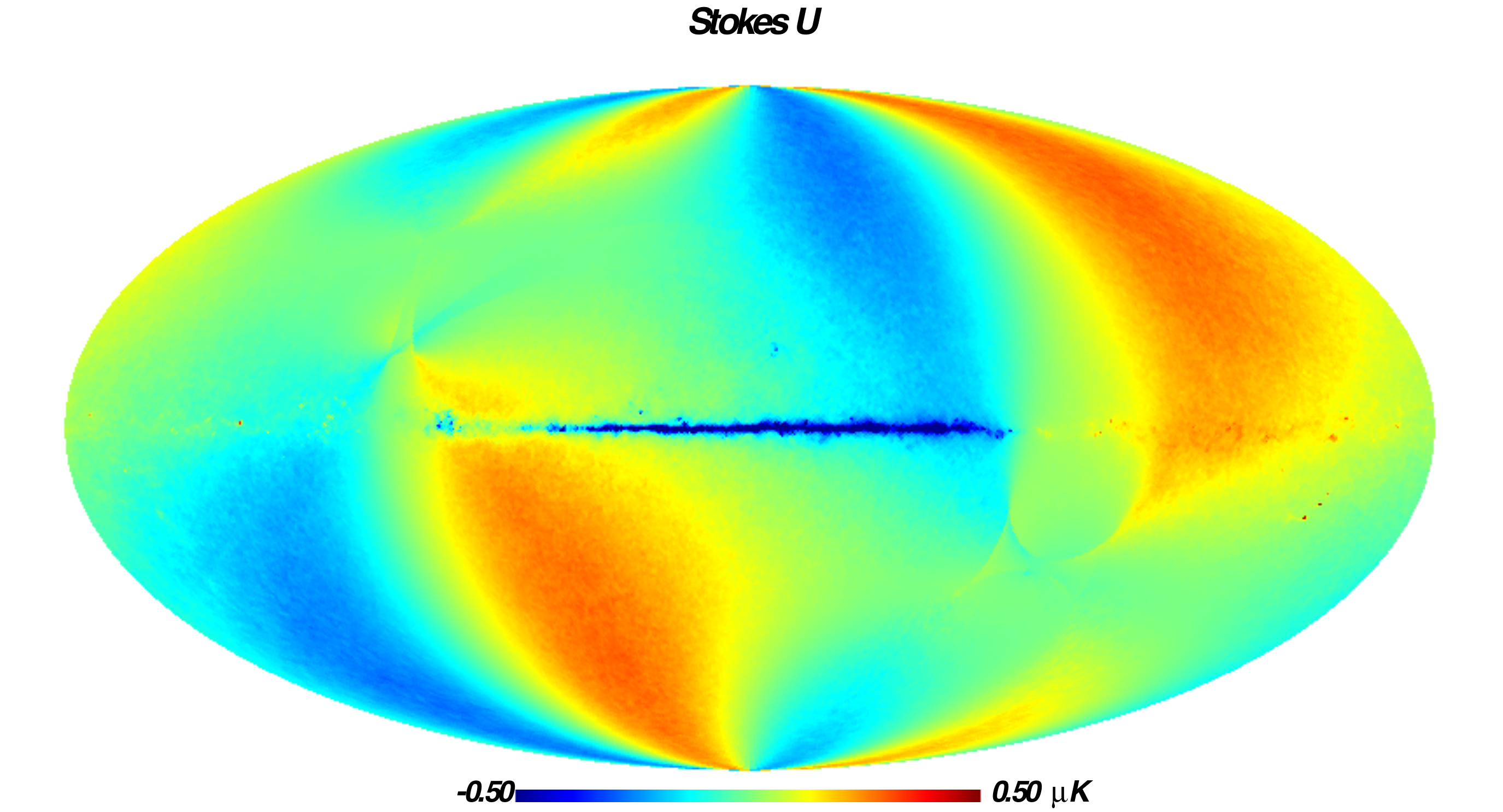}
	\caption{Residual maps of the Q and U Stokes parameters for a calibration uncertainty of $~5\times10^{-5}$. The main features on large scales are related to the CMB dipole leaking into the polarization.}
	\label{fig:residual_map_calib}
\end{figure*}

Bias in the gain reconstruction leads to a mis-estimation of the amplitude of the recovered intensity map. Its effects are however more dramatic for the polarization. If the bias affects all the detectors at the selected frequency in the same way, in the case of Planck-HFI, this would not result in a leakage of the total intensity into the polarization. However, in our case, the bias depends on the detector and it will cause a leakage of the temperature signal (dominated by the solar dipole) into $Q$ and $U$ Stokes parameters. Figure~\ref{fig:residual_map_calib} illustrates the effect of this leakage on residual $Q$ and $U$ maps for a gain error of $~5\times10^{-5}$.

\subsection{Gain statistical error}
\label{sec:calib_stat}

\begin{figure}[h!]
	\center
	\includegraphics[width=0.49\textwidth]{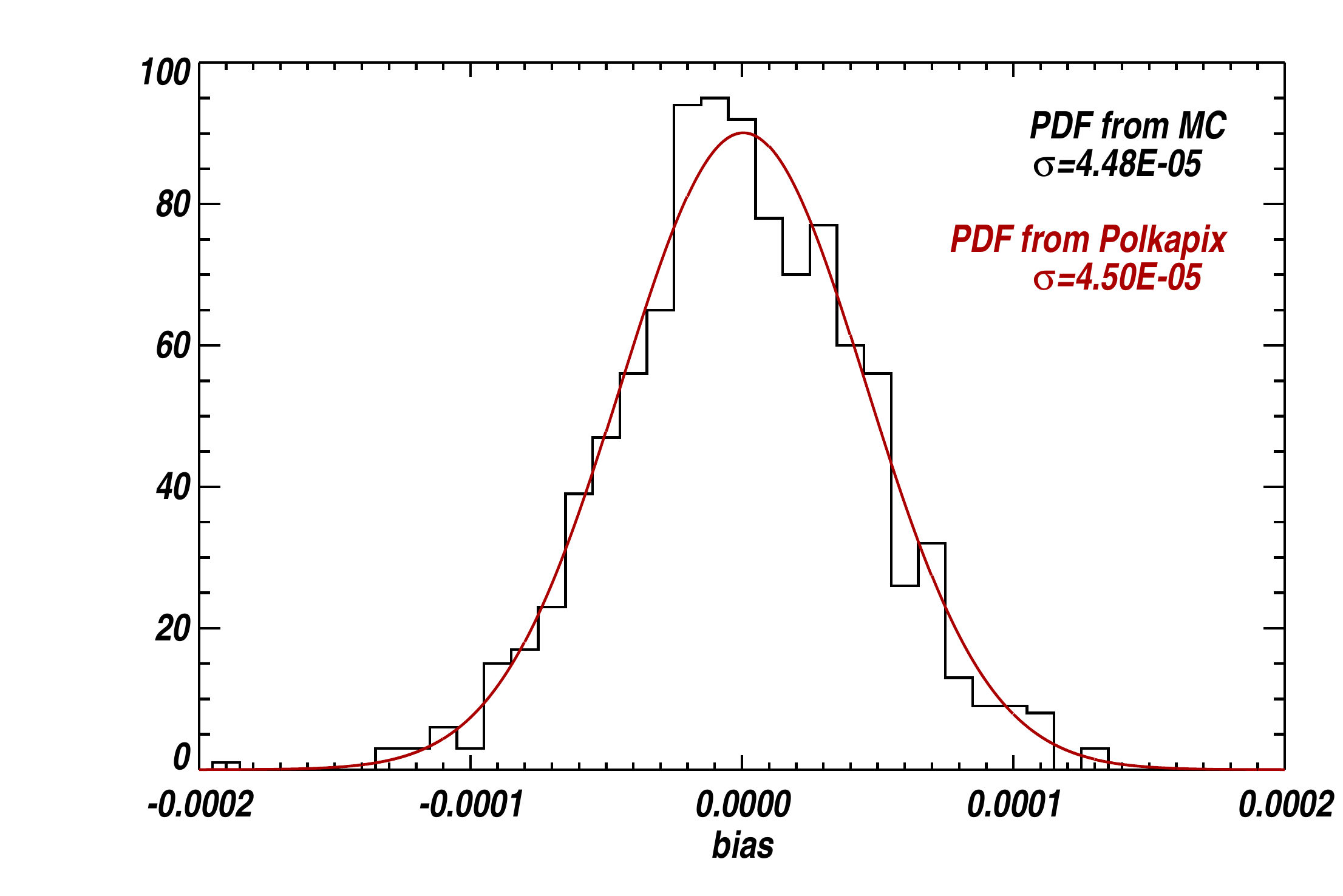}
	\caption{Gain bias distribution for 1000 simulations including CMB dipoles and pure white noise relative to {\sc Polkapix} posterior PDF. A mask covering $20$\% of the sky is applied prior to calibration.}
\label{fig:stat_calib}
\end{figure}

The uncertainty in the gain reconstruction is estimated from Eq.~\ref{eq:fisher_cal}. We have verified this estimation using $1000$ Monte Carlo simulations including only the CMB dipoles together with pure white noise for one bolometer at $143$~GHz. We apply the $20$\% Galactic mask used in the previous section. Figure~\ref{fig:stat_calib} shows the distribution of the gain bias (the relative difference between the reconstructed and simulated gains) for the 1000 simulations compared to a Gaussian with FWHM derived from Eq.~\ref{eq:fisher_cal}. The standard deviation of the MC is fully compatible with the analytical error bar given by the number of simulations. The statistical error does not depend on the input signal or noise characteristics but only on the white noise level and the orbital dipole signal. For the four bolometers at $143$~GHz, statistical errors are $\sim 4.5\times10^{-5}$.

\subsection{Calibration efficiency on the sky}
\label{sec:alphamap}

\begin{figure*}[h!]
	\center
	\includegraphics[width=0.49\textwidth]{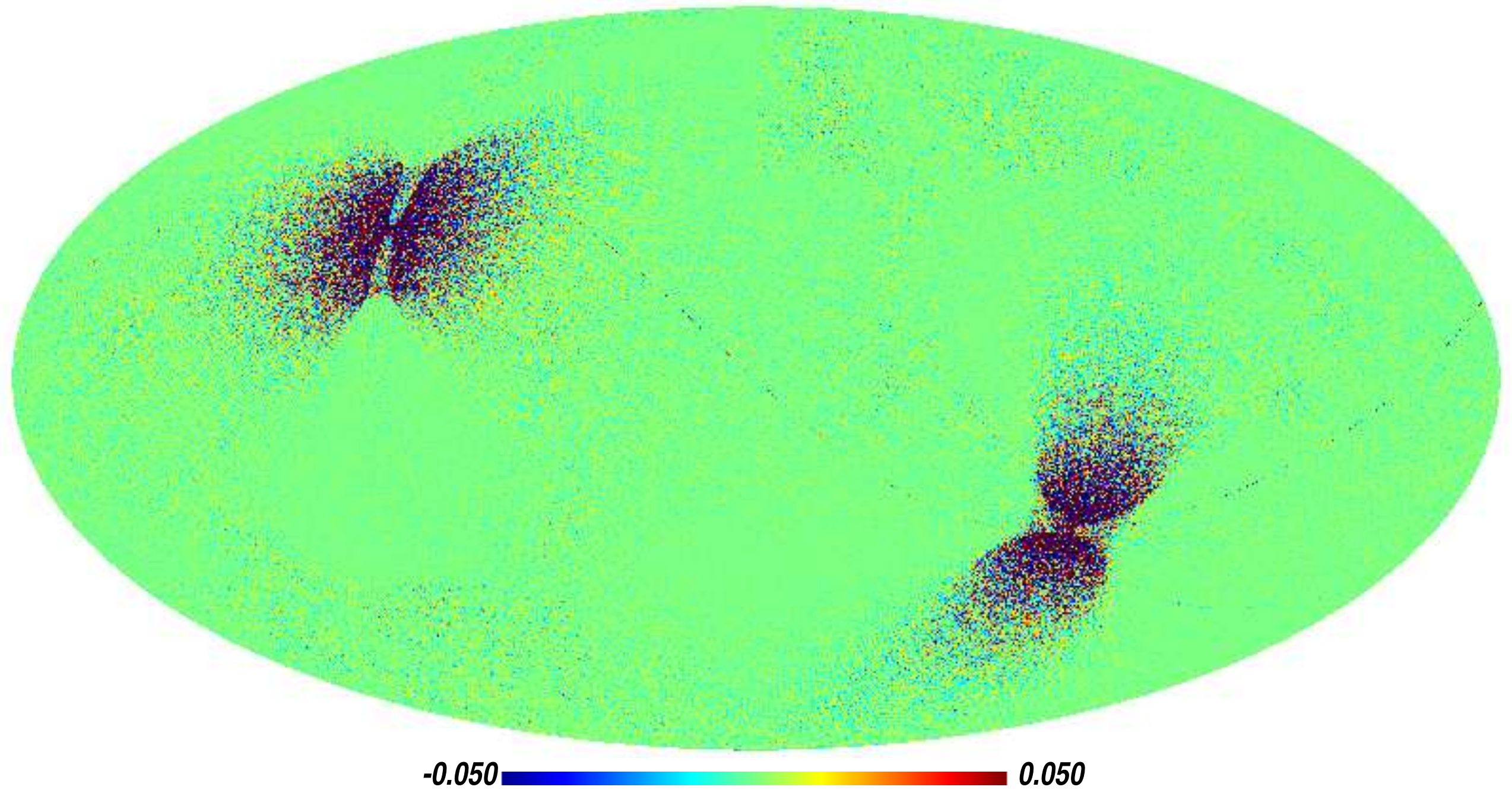}
	\includegraphics[width=0.49\textwidth]{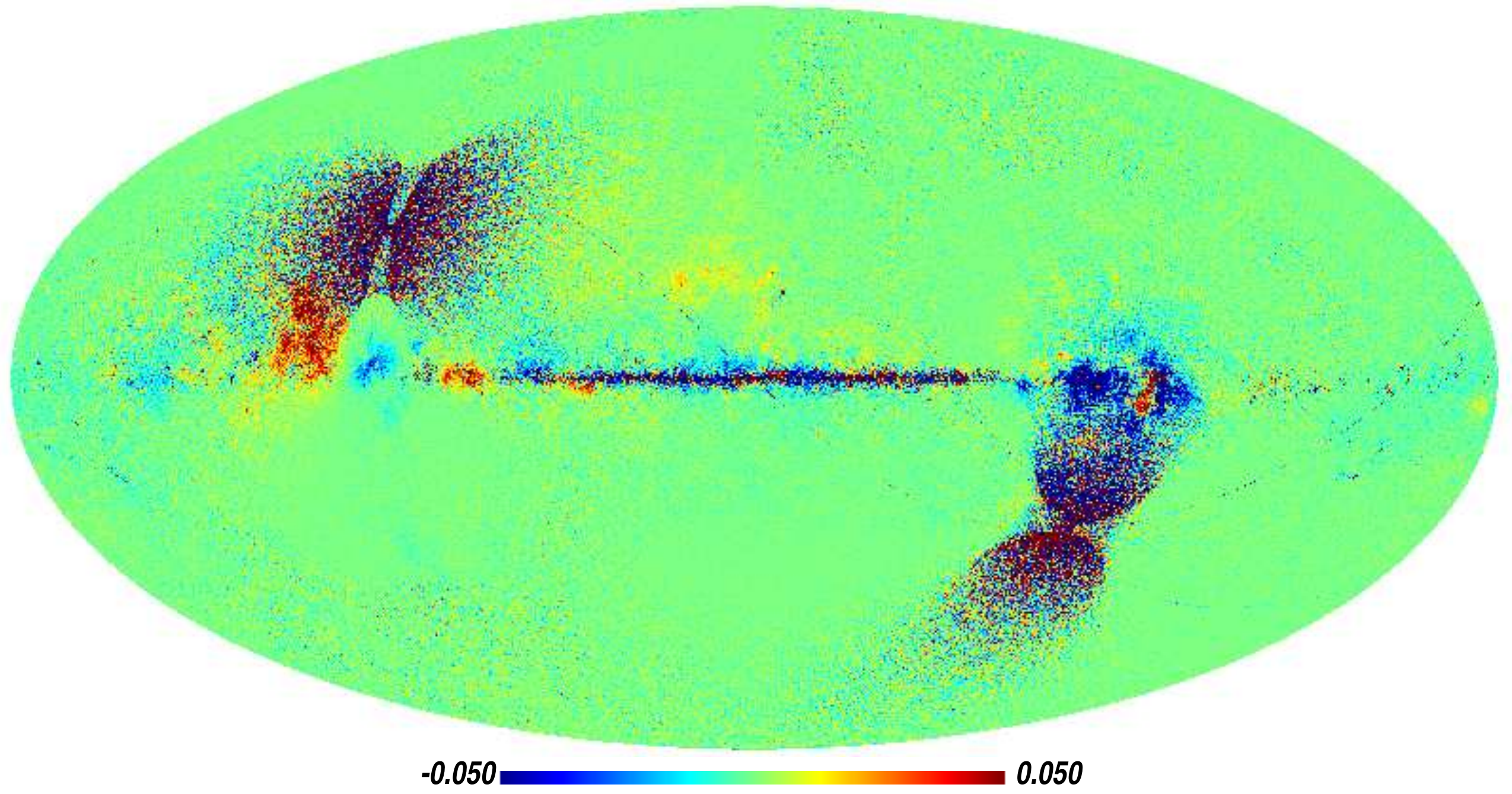}
	\caption{Maps of the contribution of each sky pixel to the global gain bias obtained for CMB-only ({\it left}) and CMB+galaxy ({\it right}) simulations.}
	\label{fig:alphamap}
\end{figure*}

\begin{figure}[h!]
	\center
	\includegraphics[width=0.49\textwidth]{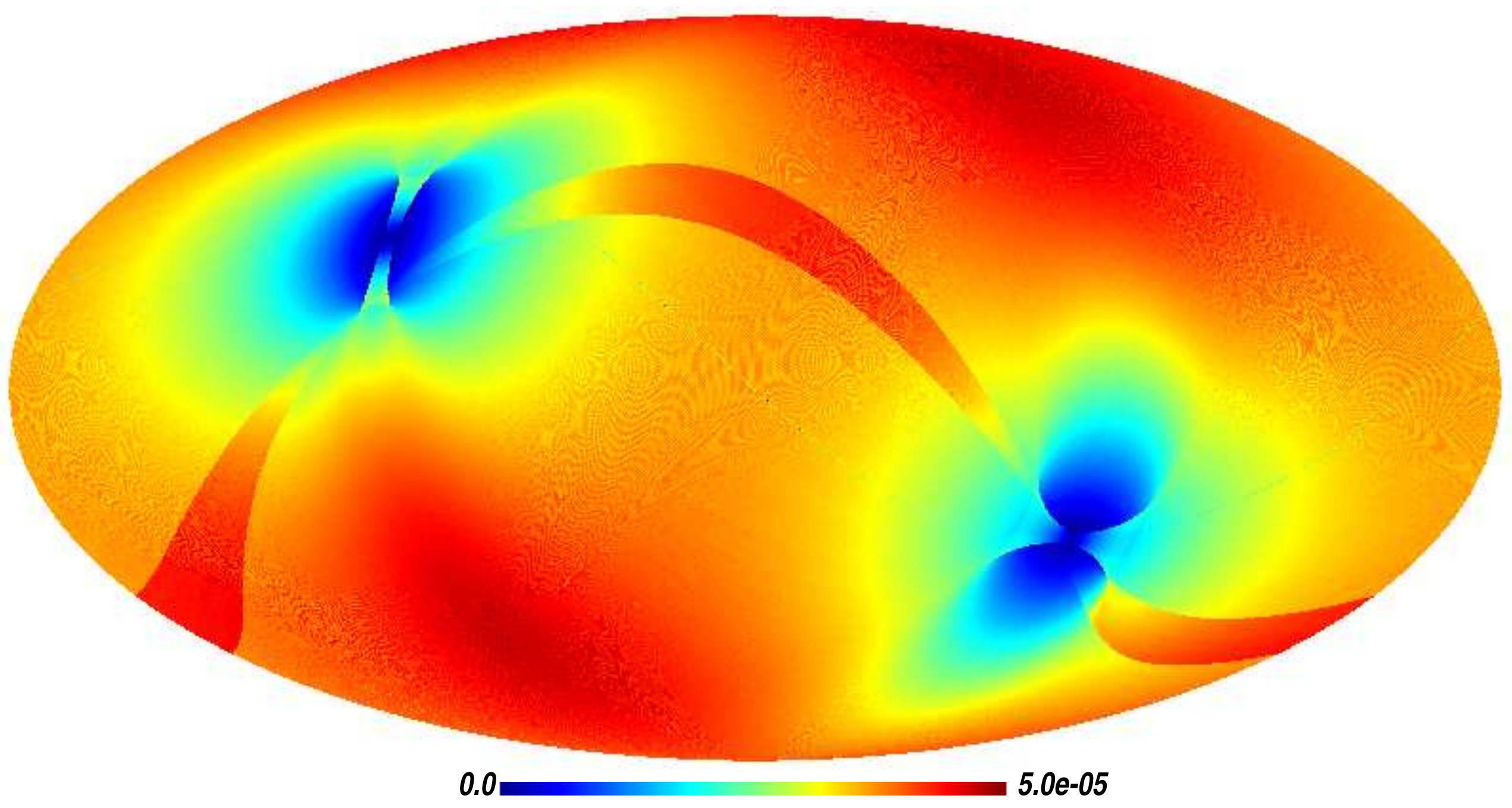}
	\caption{Map of the weight applied to the contribution per pixel toward the global gain illustrating the sensitivity to orbital dipole. The high-level sensitivity band matchs the overlap with the beginning of the third survey (see also Fig.~\ref{fig:hitcounts}).}
	\label{fig:dalphamap}
\end{figure}

From the calibration algorithm, we derive the contribution from each sky pixel to the global gain bias for a simulation including either CMB signal only or CMB and Galactic emissions (Fig.~\ref{fig:alphamap}). Around the Ecliptic poles, this contribution displays large variations because the dipole has a lower signal-to-noise ratio. In these regions, the line of sight is orthogonal to the satellite speed direction, which remains close to the Ecliptic plane, leading to a small amplitude orbital dipole signal. We note also that whenever Galactic emissions are included the contributions from the Galactic plane pixels become very significant, which given that the sky signal in this area is notoriously difficult to estimate with sufficient fidelity, could give rise to a bias, as indeed already pointed out in Fig.~\ref{fig:calib_bias_Polar}. The impact of the Galactic plane is magnified by the weights, which need to be applied to the contribution of each pixel and are shown in Fig.~\ref{fig:dalphamap}. These define the sensitivity to the orbital dipole for each pixel. In particular, we see that the sensitivity is much higher in the Galactic plane than at the Ecliptic poles, hence the former not the latter is more likely to be important to the gain estimation. This observation justifies the use of a $20$\% Galactic mask and explains the observations made in Sect.~\ref{sec:calib_syste}. 

\section{Solving for offsets and gain by iteration}
\label{sec:iteration_results}

Precise knowledge of the gains is required for destriping in order to precisely estimate the sky ($I,Q,U$) and the offsets, and remove the non-stationary component (the orbital dipole). In contrast, calibration cannot be performed before the removal of the low-frequency noise, i.e. the subtraction of the offsets from the data. This leads to the following iterative pipeline scheme. At each iteration step, we calibrate the detectors independently and then estimate offsets on a multi-detector basis. We repeat these operations until the relative difference between two consecutive gain estimations for each detector is smaller than $10^{-7}$.

\subsection{Characterization of the convergence}

\begin{figure}[h!]
	\center
	\includegraphics[width=0.49\textwidth]{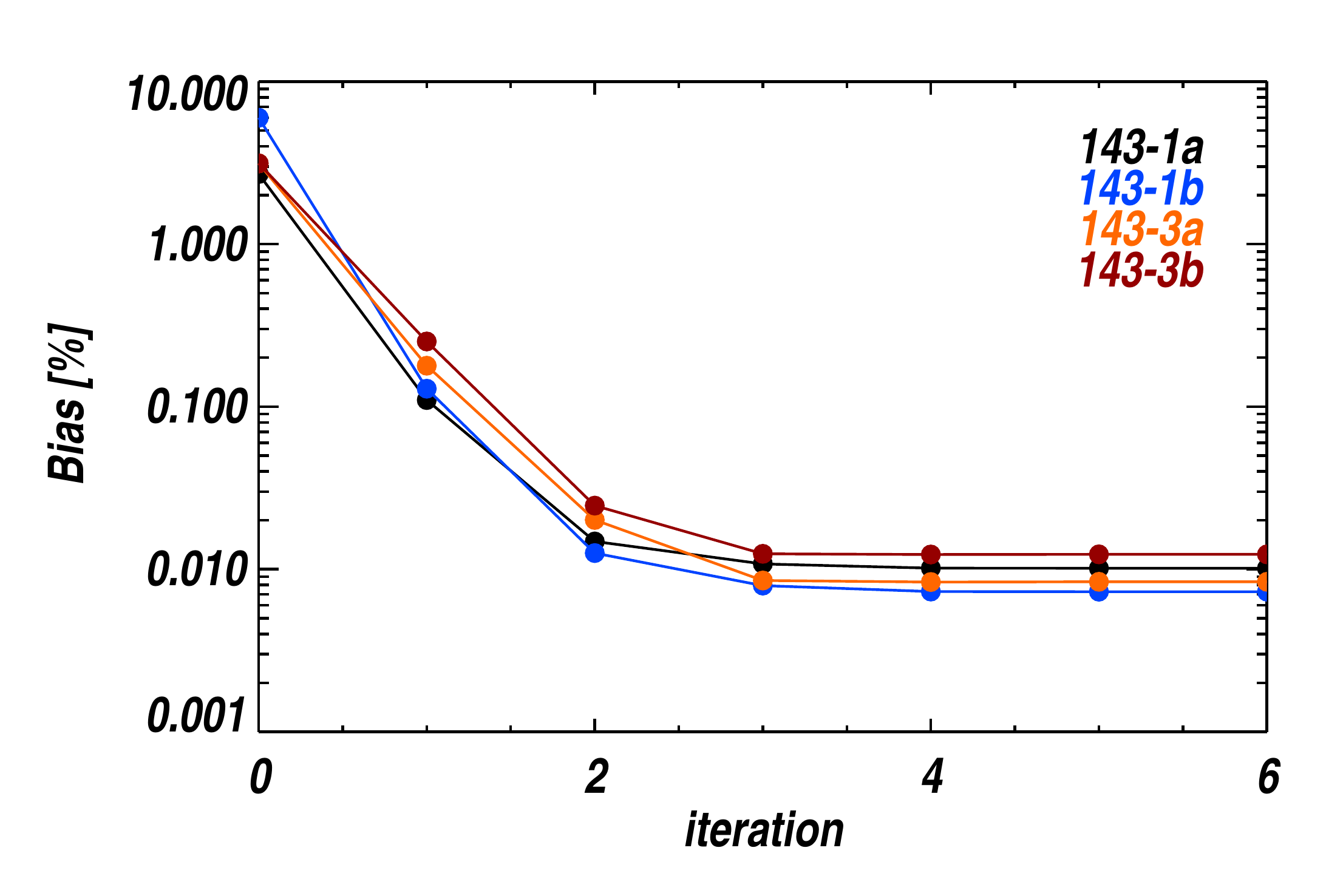}
	\caption{Bias of the reconstructed gains versus an iteration number in percent for a simulation with CMB, Galactic emission and pure white noise. The iterations are stopped when the relative differences between consecutive gains reaches $10^{-7}$.}
	\label{fig:convergence_gain}
\end{figure}

\begin{figure}[h!]
	\center
	\includegraphics[width=0.49\textwidth]{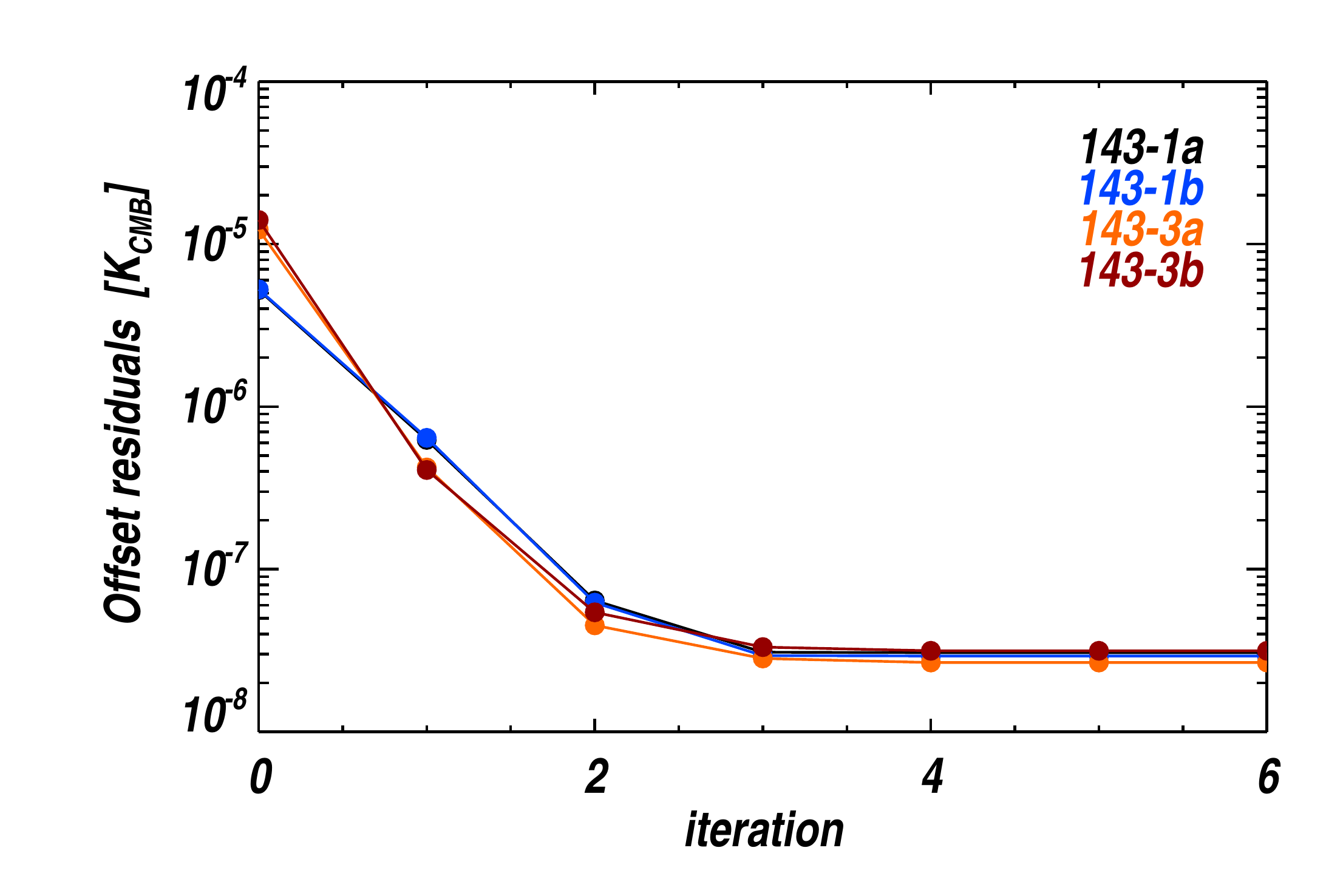}
	\caption{R.m.s. of the offset residuals with respect to iteration number in $K_\mathrm{CMB}$ for a simulation with CMB, Galactic emission, and pure white noise.}
	\label{fig:convergence_offsets}
\end{figure}

To check the convergence of the iterations for both gain and offset reconstruction, we present in Fig.~\ref{fig:convergence_gain} an example of the decrease in the gain bias with respect to the number of iterations for one of our simulations of signal (CMB and Galactic emissions) plus pure white noise. The initial condition corresponds to random gain factors generated with 5\% r.m.s. of the simulated value. We reach the convergence level in about five iterations. Figure~\ref{fig:convergence_offsets} shows, for the same simulation, the r.m.s of the offset residuals with respect to the iteration number.

\subsection{Gain and sky map accuracy} 
We now present results from the analysis of simulations including four bolometers at 143~GHz (see Sect.~\ref{sec:simulations} for details). The sky signal is the combination of CMB (including the dipoles) and Galactic emissions to which we have added different noise TOD of pure white noise, or correlated noise with either $(\alpha=2,f_{knee}$ = 0.01 or 0.1 Hz) or ($\alpha=1,f_{knee}=0.1$~Hz). One hundred noise realizations were generated in each case. Table~\ref{tab:calib} describes the results of the gain reconstruction for one bolometer (143-1a). As for the destriping (see Sect.~\ref{sec:destriping_fknee}), the calibration precision is significantly worse in this case ($\alpha=2,f_{knee} =0.1$~Hz). However, even in this case, the relative precision is on the order of a few $10^{-4}$.

\begin{table}[htdp]
	\center
	\begin{tabular}{lcccc}
		\hline\hline
		Simulation & White & $f_{knee}=0.01$ & $f_{knee}=0.1$ & $f_{knee}=0.1$ \\
		 & Noise & $\alpha=2$ & $\alpha=2$ & $\alpha=1$ \\
		\hline
gain bias & $ 8.6\times10^{-5}$ & $ 8.4\times10^{-5}$ & $ 6.8\times10^{-5}$ & $ 8.5\times10^{-5}$ \\
statistical error & $ 4.5\times10^{-5}$ & $ 4.5\times10^{-5}$ & $ 4.5\times10^{-5}$ & $ 4.5\times10^{-5}$ \\
polar systematic & $ 5.3\times10^{-5}$ & $ 5.3\times10^{-5}$ & $ 5.3\times10^{-5}$ & $ 5.3\times10^{-5}$ \\
$1/f^\alpha$ systematic & $ 0.1\times10^{-5}$ & $ 1.1\times10^{-5}$ & $25.0\times10^{-5}$ & $ 6.9\times10^{-5}$ \\
		\hline
	\end{tabular}
	\caption{Gain biases and errors for bolometer 143-1a, for the four simulated noises: pure white noise, correlated noise with $(\alpha=2,f_{knee}$ = 0.01 or 0.1 Hz), and ($\alpha=1,f_{knee}=0.1$~Hz). Statistical error is constant (see Sect.~\ref{sec:calib_stat}). Polarized signals introduce a systematic bias which we reduce to $5.3\times10^{-5}$ by removing 20\% of the brightest Galactic regions (see Sect.~\ref{sec:calib_syste}).}
\label{tab:calib}
\end{table}

\begin{figure*}[ht]
	\center
	\includegraphics[width=0.7\textwidth]{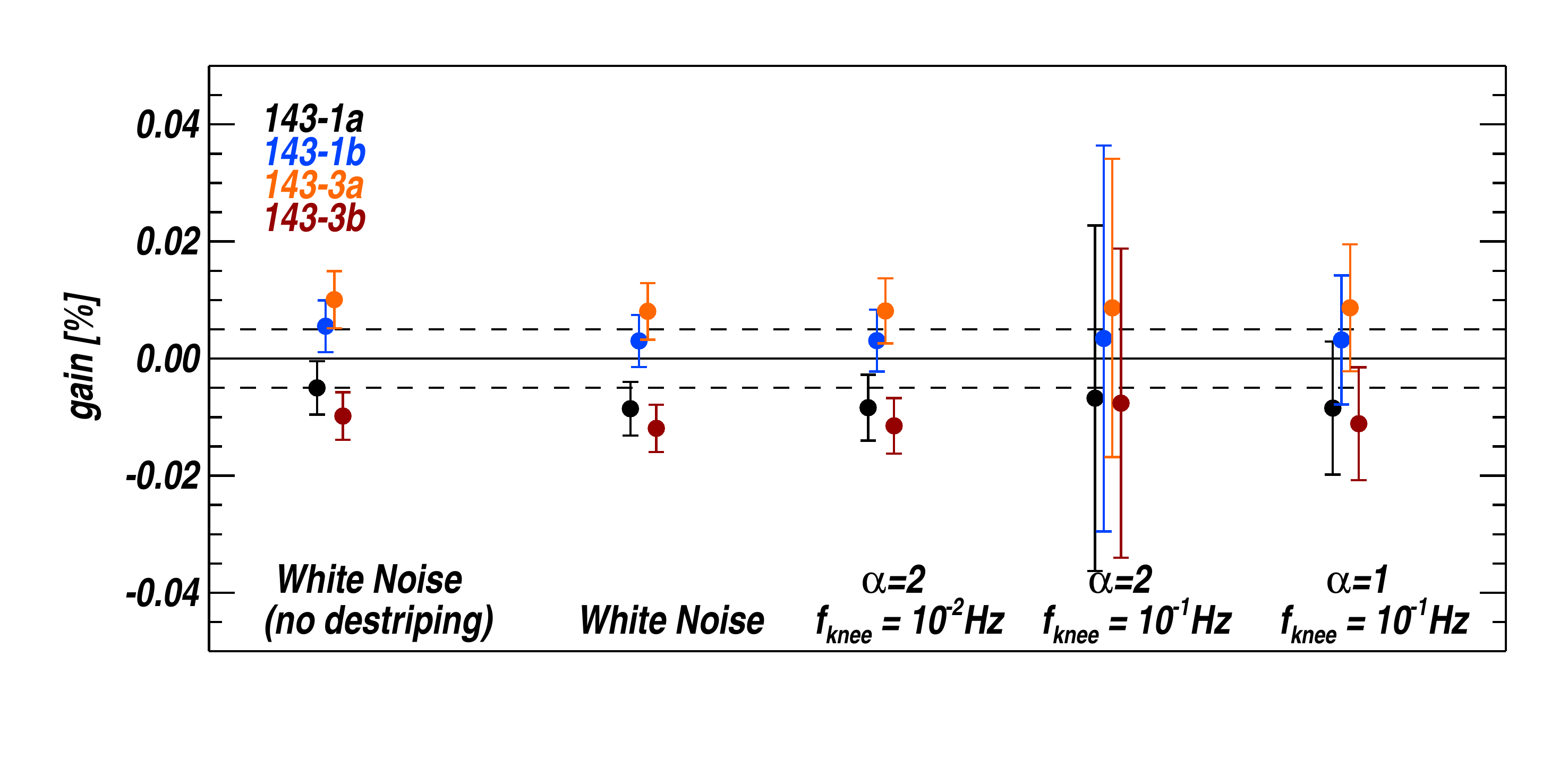}
	\caption{Gain biases for the four bolometers averaged over the Monte Carlo simulations, considering each dataset ({\it from left to right:} pure white noise without destriping, pure white noise after iterations, and low frequency noise with ($\alpha=2,f_{knee} = 0.01$), ($\alpha=2,f_{knee} = 0.1$), and ($\alpha=1,f_{knee} = 0.1$)). Error bars are derived from Monte Carlo simulations. {\it dash line: } systematics estimated in Sect.~\ref{sec:calib_syste}.}
	\label{fig:calib}
\end{figure*}

Figure~\ref{fig:calib} summarizes the calibration reconstruction accuracy for the four bolometers at 143~GHz. The gain biases are all consistent with zero given the statistical and systematic errors. The first set corresponds to gains reconstructed without destriping, for the pure white noise simulations. The following points represent the results of the pipeline for each dataset. The averaged gain for each bolometer is stable when changing the noise properties. The systematic bias, on the order of $5\times10^{-5}$, results from the anisotropies of the CMB polarization, as explained in section \ref{sec:calib_syste}.

\begin{figure*}[!h]
	\center
	\includegraphics[width=\textwidth]{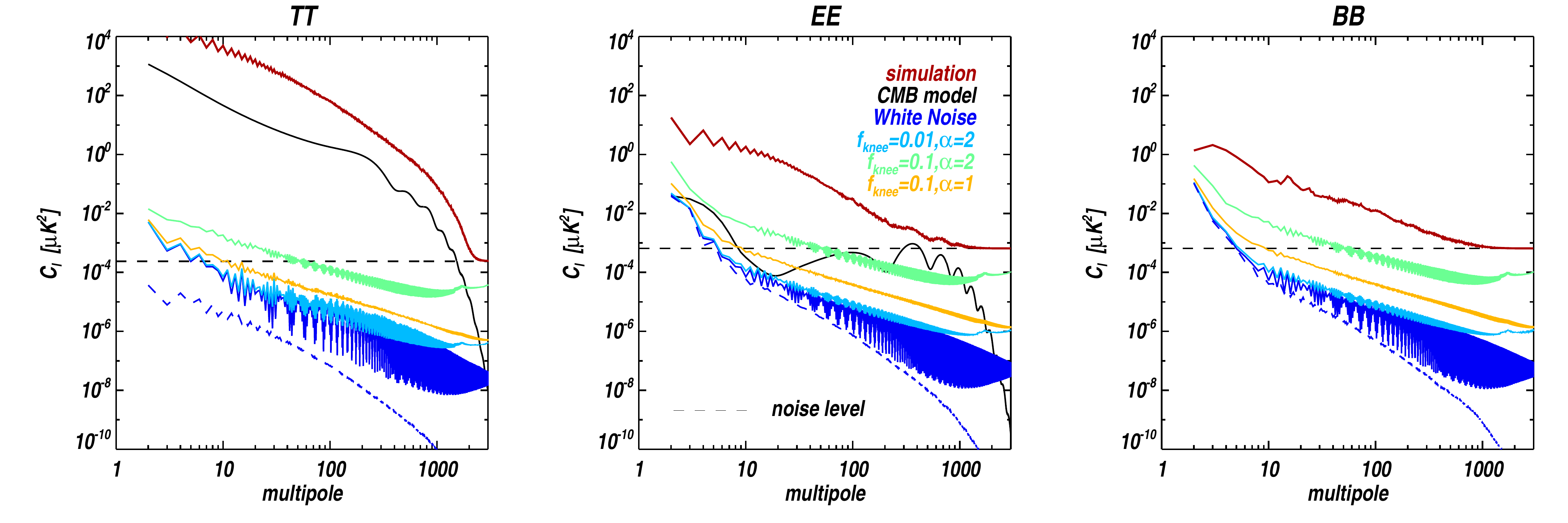}
	\includegraphics[width=\textwidth]{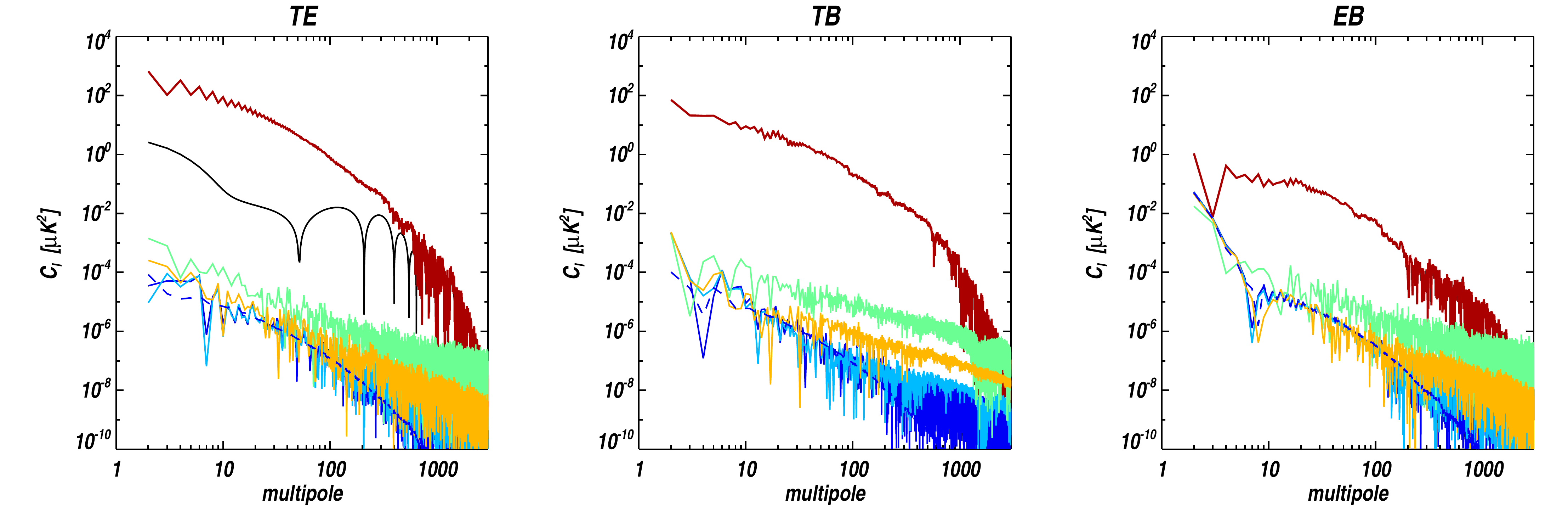}
	\caption{Average full-sky spectra of noise residual maps after the destriping and calibration pipeline for the datasets: pure white noise, and low frequency noise plus white noise with $f_{knee} = 0.1$~Hz and 0.01~Hz. The effect of the calibration uncertainty is represented by the spectra of the residual map constructed using estimated gains ({\it blue dashed line}). The simulation power spectra and CMB WMAP-5yr fiducial model are also plotted for comparison. {\it Upper line, from left to right:} temperature, E-mode, B-mode. {\it Bottom line, from left to right:} cross-correlation TE, TB, and EB.}
\label{fig:iteration_residuals}
\end{figure*}

To evaluate the quality of the sky map reconstruction from our pipeline, we used a similar procedure to that adopted for the destriping analysis presented in Sect.~\ref{sec:destriping_results}. Figure~\ref{fig:iteration_residuals} presents a compilation of averaged power spectra from residual maps for each datasets. The residual spectra have an additional excess power at low $\ell$ (below $\sim 20$) compared to those shown in Fig.~\ref{fig:fknee_pol}. We attribute this excess to the calibration uncertainty that induces a leakage, of roughly 1~$\mu K$, of the orbital dipole signal into polarization. We computed the power spectra of the difference in the pure signal $I$, $Q$, $U$ maps built with the recovered gains and the input maps (Fig.~\ref{fig:residual_map_calib}). As shown in Fig.~\ref{fig:iteration_residuals}, it matches the increase in power in the residual power spectra.

\begin{figure*}[!h]
	\center
	\includegraphics[width=0.49\textwidth]{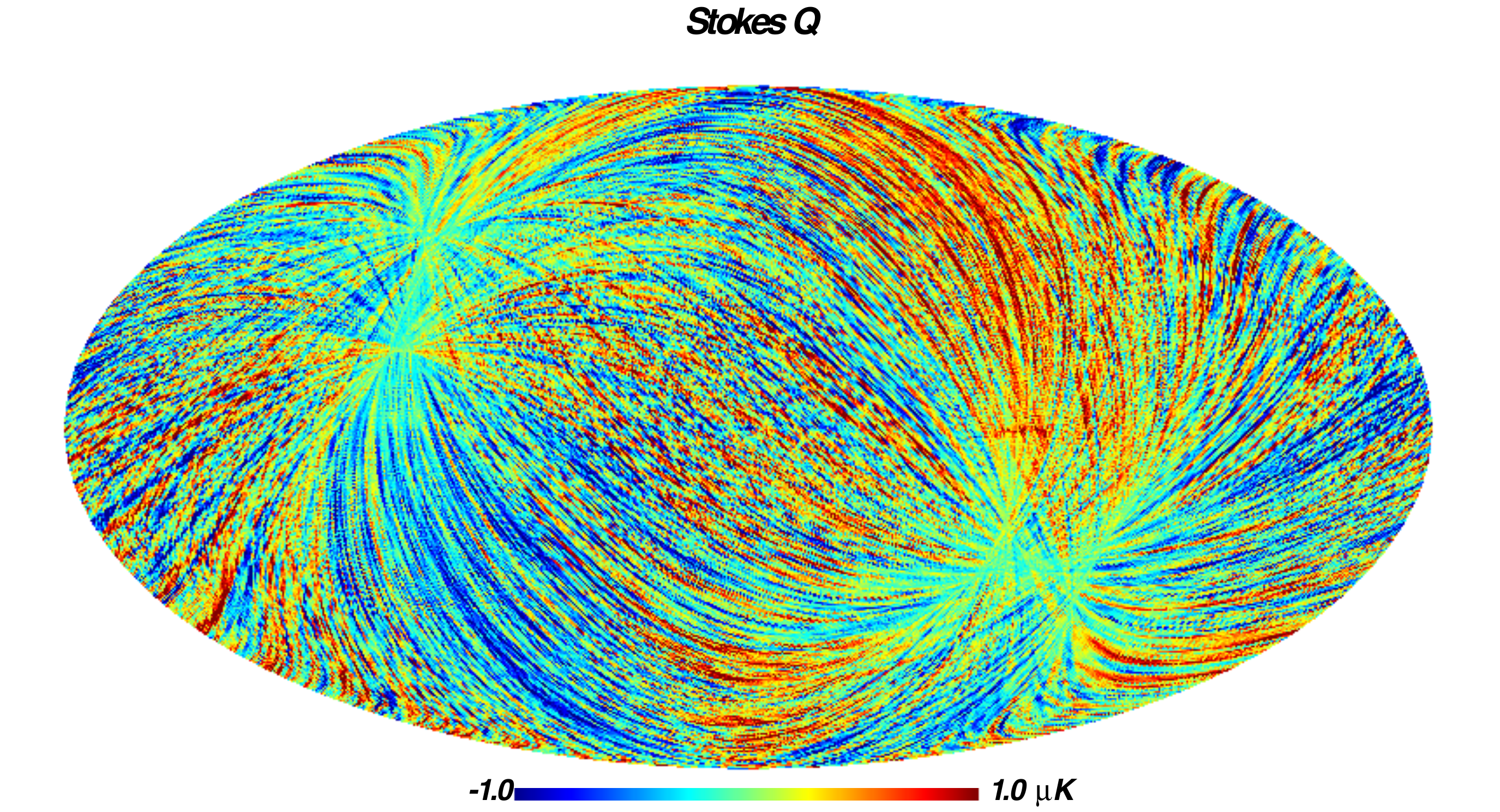}
	\includegraphics[width=0.49\textwidth]{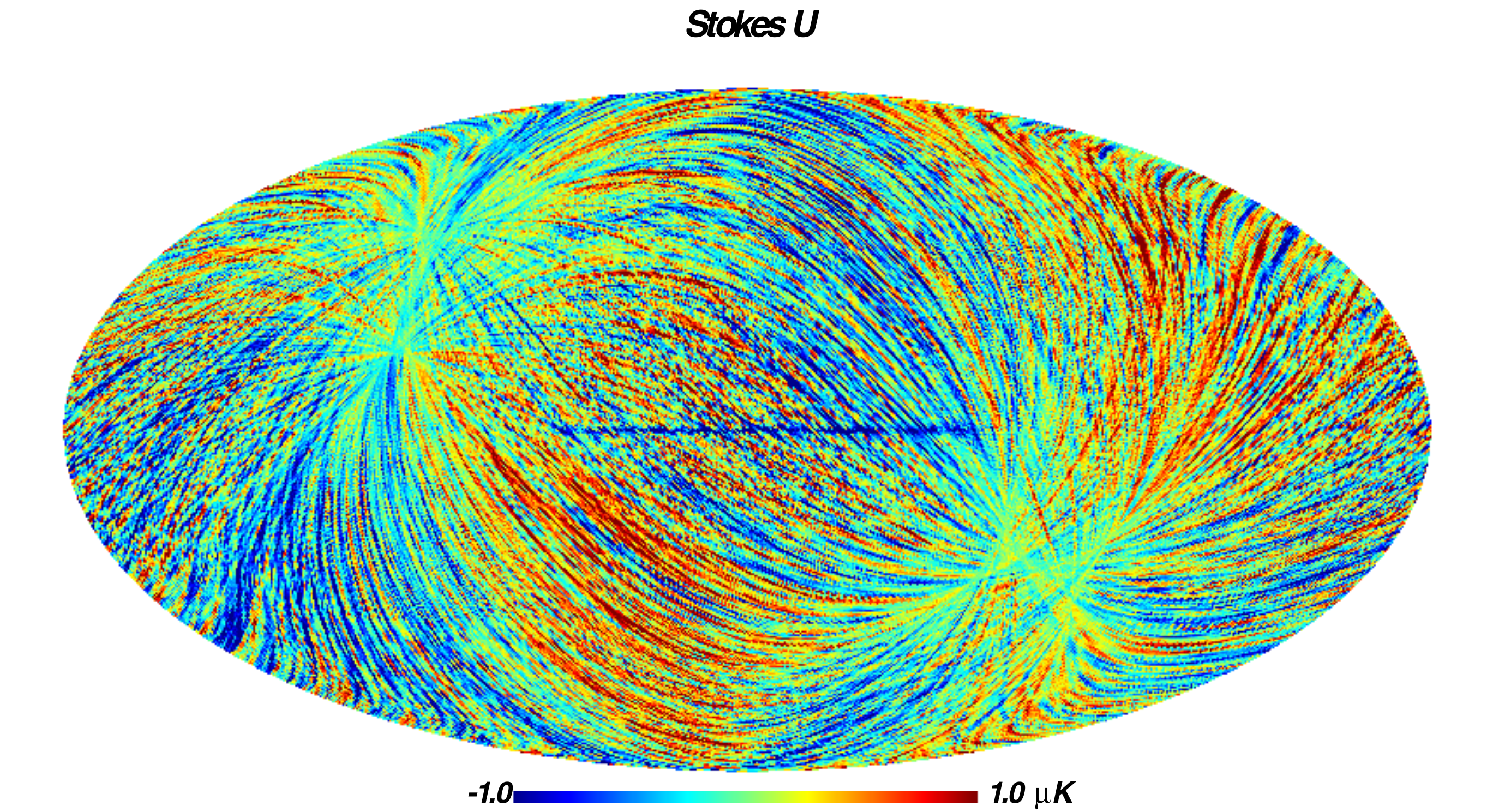}
	\caption{Residual maps of the Q and U Stokes parameters after iteration on a simulation of signal (CMB+Galactic emissions) plus low frequency noise ($\alpha=1,f_{knee} = 0.1$). Maps are degraded at 27.5~arcmin (HEALPix nside = 128). Large-scale residuals are compatible with the orbital dipole leakage as shown in Fig.~\ref{fig:residual_map_calib}. Stripes are destriping residuals.}
	\label{fig:residual_map_iteration}
\end{figure*}

Figure~\ref{fig:residual_map_iteration} shows an example of the residual maps after iterations for one of the MC simulation. We clearly see the large-scale structures correlated with the orbital dipole residuals together with the stripes produced by the destriping errors.

\section{Conclusion}

We have presented an iterative scheme for map-making, by means of the destriping and photometric calibration of {\it Planck}-HFI data. This method has been fully implemented and tested within the HFI Data Processing Center. We have used simulated datasets for 143 and 217~GHz HFI detectors, under various noise hypotheses to evaluate the performances and derive the impact of systematics.

We have first set the parameters of the destriping to minimize the systematic effects that could bias the offset determination (foregrounds signal, sky pixelization). We have shown that for correlated noise with a knee frequency up to 0.1 Hz, destriping residuals are below the white noise level except for $\ell < 50$ for temperature and polarization. We found a significant bias for polarization at a knee frequency above 0.01~Hz. For a Planck-like Fourier spectrum presented by \citet{edp}, we found residuals below the white noise level above $\ell = 5$.

We have then studied the performances of a gain reconstruction scheme based on the orbital dipole signal. We have demonstrated that in the same noise hypothesis as previously, we are able to reconstruct the photometric calibration with a statistical error of $5\times10^{-5}$. For pure white noise, we have also evaluated the systematic uncertainty to be $5\times10^{-5}$, which is dominated by CMB polarization anisotropies that are not modeled in our single-bolometer calibration scheme.

Finally, we presented the results when solving for offsets and gain by iteration. With low frequency noise, the gain bias stays constant but the uncertainties increase up to a few $10^{-4}$ for correlated noise with ($\alpha=2,f_{knee}=0.1$~Hz). The calibration uncertainty induces a further worsening of the destriping residuals concentrated at very low multipoles ($\ell<20$).
Altogether, this scheme could be fruitfully applied to the {\it Planck}-HFI flight data, provided they satisfy our very basic noise hypotheses.

\begin{acknowledgement}
We acknowledge the use of the Planck Sky Model, developed by the Component Separation Working Group (WG2) of the Planck Collaboration. We thank also the CTP working group for fruitful discussions on map-making algorithms. We acknowledge use of CC-IN2P3 facilities. The authors would like to thank F. Couchot for a detailed review of our manuscript.
\end{acknowledgement}


\end{document}